\documentclass{ws-ijmpd}
\usepackage[super,compress]{cite}

\usepackage{hyperref,amsfonts,amsmath,amssymb,graphicx,bm,subfigure}

\numberwithin{equation}{section}

\begin{document}

\markboth{Maxim Dvornikov}
{Impact of hypermagnetic fields on gravitational waves, neutrinos and baryon asymmetry}

%
\catchline{}{}{}{}{}
%

\title{IMPACT OF HYPERMAGNETIC FIELDS ON RELIC GRAVITATIONAL WAVES, NEUTRINO
OSCILLATIONS AND BARYON ASYMMETRY}

\author{MAXIM DVORNIKOV}

\address{Pushkov Institute of Terrestrial Magnetism, Ionosphere \\ and Radiowave Propagation (IZMIRAN), \\
108840 Moscow, Troitsk, Russia\\
maxdvo@izmiran.ru}

\maketitle


\begin{abstract}
We study the evolution of random hypermagnetic fields (HMFs) in the
symmetric phase of the early universe before the electroweak phase
transition. The behavior of HMFs is driven by the analog of the chiral
magnetic effect accounting for the asymmetries of leptons and Higgs
bosons. These asymmetries are also dynamical variables of the model and evolve 
together with HMFs. Moreover, we account for the contribution of the hyper-MHD
turbulence in the effective diffusion coefficient and the $\alpha$-dynamo
parameter. The realistic spectrum of seed HMFs consists of two branches:
Batchelor and Kolmogorov ones. The impact of HMFs on the production
of relic gravitational waves (GWs) and the baryon asymmetry of the
universe (BAU), as well as flavor oscillations of supernova neutrinos
in the stochastic GWs generated, are considered. We establish the constraint on
the strength of the seed HMF comparing the spectral density of produced
GWs with the observations of the LIGO-Virgo-KAGRA collaborations.
The stronger upper bound on the seed HMF is obtained from the condition
of not exceeding the observed value of BAU.
\end{abstract}



\section{Introduction}

The origin of cosmic magnetic fields is a puzzle for the modern astrophysics,
cosmology, and particle physics. If one believes in the existence
of nonzero magnetic fields in intergalactic voids, as suggested in
Ref.~\refcite{NerVov10}, it is difficult to explain the production
of such fields by astrophysical means. Thus large scale cosmic magnetic
fields are likely to be of cosmological origin~\cite{DurNer13}.
Cosmological magnetic fields with proper characteristics can be generated, e.g., 
during inflation, in the QCD and electroweak phase transitions (EWPT)
and due to the Higgs field gradient. Some of the mechanisms for the
production of such magnetic fields are reviewed in Ref.~\refcite{Sub16}.

Maxwell cosmological magnetic fields can stem from hypermagnetic fields
(HMFs) which exist before EWPT. These HMFs result from the massless
hypercharge field $Y^{\mu}$ which is present in the symmetric phase
of the universe evolution. A hypercharge field is a linear combination of electromagnetic and $Z$-boson fields. The connection between the production of
HMFs and the leptogenesis, as well as the baryogenesis was suggested
in Ref.~\refcite{JoySha97}. It is possible owing to the abelian anomaly
for the hypercharge field. Thus, a configuration of HMFs decays and
creates leptons which are massless particles before EWPT. This scenario
was further developed in Refs.~\refcite{DvoSem13,KamLon16}.

On the other hand, there is a backreaction from the lepton asymmetries
to the evolution of (hyper)-magnetic fields. It is based on the chiral
magnetic effect (the CME)~\cite{Fuk08}, which consists in the modification
of the induction equation in the presence of a chiral imbalance, or
an asymmetry, of ultrarelativistic fermions. A magnetic field becomes
unstable in this case. The application of the CME is justified in
the symmetric phase. However, one can use it in the broken phase as
well~\cite{BoyFroRuc12} by accounting for the appropriate spin flip
rate~\cite{BoyCheRucSob21}.

Besides the production of the baryon asymmetry of the universe (BAU),
cosmological magnetic fields, which have a random structure, can result
in the generation of relic gravitational waves (GWs). This problem
was studied, e.g., in Ref.~\refcite{KosMacKah02}. The production of
GWs in the primordial chiral plasma accounting for the CME was analyzed
in Ref.~\refcite{Bra21}. The recent interest to the studies of GW backgrounds
of the cosmological origin~\cite{CapFig18} is inspired by the direct
detection of GWs~\cite{Abb16}, as well as the claims (see, e.g.,
Ref.~\refcite{Arz20}) that stochastic GWs can be achievable with
modern GW detection techniques.

HMFs can indirectly influence neutrino oscillations. We found in Refs.~\refcite{Dvo19,Dvo20,Dvo21}
that neutrino flavor oscillations can be affected by stochastic GWs.
Suppose that relic GWs are produced by HMFs, as described above. Then,
these GWs interact with astrophysical neutrinos modifying the oscillations
picture and changing their fluxes. Thus, we can say that the fluxes
of astrophysical neutrinos are influenced by HMFs. Neutrino flavor
oscillations in GWs were also considered in Ref.~\refcite{KouMet19}.
The research on the interaction between neutrinos and GWs is inspired
by various multimessenger studies~\cite{Alb19,Aar20a}, where both neutrinos
and GWs are explored. Of course, existing neutrino telescopes are always waiting for a nearby supernova (SN)~\cite{VitTamRaf20}, as well as trying 
to detect a SN neutrino background. It allows one to study the interaction between neutrinos and GWs.

This work is organized in the following way. We start in Sec.~\ref{sec:EVOLHMF}
with the formulation of the basic equations for the evolution of random
HMFs. In Sec.~\ref{subsec:SEEDSPEC}, we set up the initial condition.
We present the numerical solution of the evolution equations for the
spectra and the asymmetries in Sec.~\ref{subsec:NUMHMF}. We study
the generation of primordial GWs by HMFs in Sec.~\ref{sec:PRODGWs}.
The results of the GWs production are represented in Sec.~\ref{subsec:RESGWs}.
We also consider the possibility to observe the predicted GW background
in Sec.~\ref{subsec:RESGWs}. Neutrino flavor oscillations under
the influence of relic GWs, generated in frames of our model, are
discussed in Sec.~\ref{sec:NUFLOSC}. Finally, we study the production
of BAU by the evolving asymmetries of particles in Sec.~\ref{sec:BAU}. In Sec.~\ref{sec:CONCL} we conclude.
In~\ref{sec:MHDTURB}, we rederive the contribution of the
(H)MHD turbulence to the effective diffusion coefficient and the $\alpha$-dynamo
parameter. We introduce the new variables for numerical simulations
of the HMFs evolution in~\ref{sec:NEWVAR}. The expression for the function characterizing the energy spectrum of stochastic GWs is derived in~\ref{sec:OMEGDER}. In~\ref{sec:DENSMATR},
we clarify some of the issues in the derivation of the density matrix
equation for flavor neutrinos interacting with stochastic GWs.

\section{Evolution of HMFs\label{sec:EVOLHMF}}

The key issue in our study is the evolution of HMFs, as well as the
lepton and Higgs boson asymmetries. This problem was considered in details
in Refs.~\refcite{DvoSem21,Dvo22}.

We study the situation when HMFs evolve before EWPT. The behavior
of HMFs accounts for the instability in the presence of nonzero particle
asymmetries. The full set of the kinetic equations has the form~\cite{DvoSem21,Dvo22},
\begin{align}\label{eq:HMFsys}
  \frac{\partial\tilde{\mathcal{E}}_{{\rm B_{\mathrm{Y}}}}}{\partial\tilde{\eta}} & =
  -2\tilde{k}^{2}\eta_{\mathrm{eff}}\tilde{\mathcal{E}}_{{\rm B_{\mathrm{Y}}}}+
  \alpha_{\mathrm{eff}}\tilde{k}^{2}\tilde{\mathcal{H}}_{{\rm B_{\mathrm{Y}}}},
  \nonumber
  \displaybreak[2]
  \\
  \frac{\partial\tilde{\mathcal{H}}_{{\rm B_{\mathrm{Y}}}}}{\partial\tilde{\eta}} & =
  -2\tilde{k}^{2}\eta_{\mathrm{eff}}\tilde{\mathcal{H}}_{{\rm B_{\mathrm{Y}}}}+
  4\alpha_{\mathrm{eff}}\tilde{\mathcal{E}}_{{\rm B_{\mathrm{Y}}}},
  \nonumber
  \displaybreak[2]
  \\
  \frac{\mathrm{d}\xi_{e\mathrm{R}}}{\mathrm{d}\tilde{\eta}} & =
  -\frac{3\alpha'}{\pi}\int\mathrm{d}\tilde{k}\frac{\partial\tilde{\mathcal{H}}_{{\rm B_{\mathrm{Y}}}}}{\partial\tilde{\eta}}-
  \Gamma(\xi_{e\mathrm{R}}-\xi_{e\mathrm{L}}+\xi_{0}),
  \nonumber
  \displaybreak[2]
  \\
  \frac{\mathrm{d}\xi_{e\mathrm{L}}}{\mathrm{d}\tilde{\eta}} & =
  \frac{3\alpha'}{4\pi}\int\mathrm{d}\tilde{k}\frac{\partial\tilde{\mathcal{H}}_{{\rm B_{\mathrm{Y}}}}}{\partial\tilde{\eta}}-
  \frac{\Gamma}{2}(\xi_{e\mathrm{L}}-\xi_{e\mathrm{R}}-\xi_{0})-\frac{\Gamma_{\mathrm{sph}}}{2}\xi_{e\mathrm{L}},
  \nonumber
  \displaybreak[2]
  \\
  \frac{\mathrm{d}\xi_{0}}{\mathrm{d}\tilde{\eta}} & =
  -\frac{\Gamma}{2}(\xi_{e\mathrm{R}}+\xi_{0}-\xi_{e\mathrm{L}}),
\end{align}
where $\tilde{\mathcal{E}}_{{\rm B_{\mathrm{Y}}}}(\tilde{k},\tilde{\eta})$
and $\tilde{\mathcal{H}}_{{\rm B_{\mathrm{Y}}}}(\tilde{k},\tilde{\eta})$
are the dimensionless spectral densities of the HMF energy and the helicity.
The total densities of the energy and the helicity can be computed
as $\tilde{B}_{\mathrm{Y}}^{2}/2=\smallint\mathrm{d}\tilde{k}\tilde{\mathcal{E}}_{{\rm B_{\mathrm{Y}}}}(\tilde{k},\tilde{\eta})$
and $\tilde{h}\equiv\smallint\mathrm{d}^{3}x(\tilde{\mathbf{Y}}\tilde{\mathbf{B}}_{\mathrm{Y}})/V=\smallint\mathrm{d}\tilde{k}\tilde{\mathcal{H}}_{{\rm B_{\mathrm{Y}}}}(\tilde{k},\tilde{\eta})$,
where $\tilde{\mathbf{Y}}$ is the hypercharge field in conformal
variables and $\tilde{\mathbf{B}}_{\mathrm{Y}}=(\nabla\times\tilde{\mathbf{Y}})$
is HMF. The dimensionless conformal time is $\tilde{\eta}=\tilde{M}_{\mathrm{Pl}}(T^{-1}-T_{\mathrm{RL}}^{-1})$
and the conformal momentum is $\tilde{k}=k_{\mathrm{phys}}/T$, where
$\tilde{M}_{\mathrm{Pl}}=M_{\mathrm{Pl}}/1.66\sqrt{g_{*}}$, $T$
is the primeval plasma temperature, $T_{\mathrm{RL}}=10\,\text{TeV}$ is the
temperature corresponding to the start of the evolution (see below),
$M_{\mathrm{Pl}}=1.2\times10^{19}\,\text{GeV}$ is the Planck mass,
$g_{*}=106.75$ is the number of the relativistic degrees of freedom
before EWPT, and $k_{\mathrm{phys}}$ is the physical momentum.

Along with HMFs in Eq.~(\ref{eq:HMFsys}), we account for the evolution
of the asymmetries of right and left fermions $\xi_{e\mathrm{R,L}}=6(n_{e\mathrm{R,L}}-n_{\bar{e}\mathrm{R,L}})/T^{3}$,
as well as that of Higgs bosons $\xi_{0}=3(n_{\varphi_{0}}-n_{\bar{\varphi}_{0}})/T^{3}$,
where $n_{(e,\bar{e})(\mathrm{R,L)}}$ are the number densities of
right electrons, left fermions, their antiparticles, and $n_{\varphi_{0},\bar{\varphi}_{0}}$
are the number densities of Higgs bosons and antibosons. As shown
in Ref.~\refcite{Cam92}, only the lightest lepton generation should
be taken into account since other leptons are out of equilibrium sooner
because their Yukawa coupling constants are greater. We demonstrated
in Ref.~\refcite{DvoSem13} that left fermions are to be taken into
account to transform Eq.~(\ref{eq:HMFsys}) to the closed form.

Following Ref.~\refcite{GioSha98}, in Eq.~\eqref{eq:HMFsys}, we assume that the global equilibrium in plasma before EWPT is characterized by five chemical potentials: $\mu_\mathrm{Y}$ for the conserved hypercharge, $\mu_{e\mathrm{R}}$ for right electrons, and three chemical potentials corresponding to three generations in the standard model. In Eq.~\eqref{eq:HMFsys}, additionally we account for two chemical potentials of left leptons and Higgs boson to make our analysis self-consistent since we take into account the sphaleron processes. Our approach is different from that used in Ref.~\refcite{KamLon16} where the evolution of all quarks asymmetries in the presence of HMFs was explicitly accounted for. In Refs.~\refcite{DvoSem21,SemSmiSok16}, the evolution of HMFs and the BAU generation were studied on the basis of the approach proposed in Ref.~\refcite{GioSha98}.

In Eq.~(\ref{eq:HMFsys}), the rate $\Gamma$ is caused by the interaction
of fermions with Higgs bosons. It was obtained in Ref.~\refcite{Cam92},
\begin{equation}\label{eq:spinfliprate}
  \Gamma(\tilde{\eta})=\frac{242}{\tilde{\eta}_{\mathrm{EW}}}
  \left[
    1-\frac{\tilde{\eta}^{2}}{\tilde{\eta}_{\mathrm{EW}}^{2}}
  \right],
  \quad
  \tilde{\eta}_{\mathrm{EW}}=\frac{\tilde{M}_{\mathrm{Pl}}}{T_{\mathrm{EW}}}=7\times10^{15},
\end{equation}
where $T_{\mathrm{EW}}=10^{2}\,\text{GeV}$ is the temperature of
EWPT. The dimensionless transitions rate due to the sphaleron processes
can be taken as~\cite{GorRub11} $\Gamma_{\mathrm{sph}}=8\times10^{-7}$.

We suppose that the HMFs evolution starts at $T_{\mathrm{RL}}=10\,\text{TeV}$.
This choice is justified by the fact that Higgs bosons decays become faster
than the universe expansion below this temperature. Thus, the production
of left fermions begins at $T<T_{\mathrm{RL}}$. The maximal wave
vector $\tilde{k}_{\mathrm{max}}$, which is in the integration limits
in Eq.~(\ref{eq:HMFsys}), is related to the minimal length scale.
It is the free parameter in our model. The strongest constraint on $\tilde{k}_{\mathrm{max}}$
results from the fact that the minimal scale should to be greater
than the Debye length to guarantee the plasma electroneutrality. In
various parts of our work, we shall vary $\tilde{k}_{\mathrm{max}}$
in a quite broad range: $10^{-10}<\tilde{k}_{\mathrm{max}}<10^{-2}$.
In this situation, the minimal length scale of HMFs is still greater
than the conformal Debye length $\tilde{r}_{\mathrm{D}}=10$.

The effective magnetic diffusion coefficient $\eta_{\mathrm{eff}}$
and the effective $\alpha$-dynamo parameter $\alpha_{\mathrm{eff}}$
account for the analogs of both the CME and the (H)MHD turbulence
for HMFs. They are~\cite{Cam07} (see also~\ref{sec:MHDTURB}),
\begin{align}\label{eq:etaalphaY}
  \eta_{\mathrm{eff}} & =\sigma_{c}^{-1}+
  \frac{4}{3}\frac{(\alpha')^{-2}}{\tilde{\rho}+\tilde{p}}
  \int\mathrm{d}\tilde{k}\mathcal{\tilde{E}}_{{\rm B_{\mathrm{Y}}}},
  \quad
  \alpha'=\frac{g'^{2}}{4\pi},
  \nonumber
  \\
  \alpha_{\mathrm{eff}} & =\alpha_{\mathrm{Y}}(\tilde{\eta})+
  \frac{2}{3}\frac{(\alpha')^{-2}}{\tilde{\rho}+\tilde{p}}
  \int\mathrm{d}\tilde{k}\tilde{k}^{2}\mathcal{\tilde{H}}_{{\rm B_{\mathrm{Y}}}},
\end{align}
where $\sigma_{c}\approx10^{2}$ is the conformal conductivity of
a relativistic plasma, $\tilde{\rho}$ and $\tilde{p}$ are the plasma
density and the pressure expressed in conformal variables, $g'=e/\cos\theta_{\mathrm{W}}$
is the hypercharge, and $\theta_{\mathrm{W}}$ is the Weinberg angle. We choose $p=\rho/3$ for the ultrarelativistic
plasma. The value of $\alpha'$ is $\alpha'=9.5\times10^{-3}$~\cite{DvoSem21}.

Note that the form of $\alpha_{\mathrm{eff}}$ in Eq.~\eqref{eq:etaalphaY}
is different from that in Refs.~\refcite{DvoSem21,Dvo22,DvoSem17}.
To resolve the contradiction between the results of Ref.~\refcite{Cam07}
and Ref.~\refcite{DvoSem17} we rederive the contribution of turbulent
(H)MFs to the kinetic Eq.~(\ref{eq:HMFsys}) in~\ref{sec:MHDTURB};
cf. Eq.~(\ref{eq:etaalpha}). We confirm the validity of $\alpha_{\mathrm{eff}}$
in Ref.~\refcite{Cam07}. However, as mentioned in Refs.~\refcite{DvoSem21,Dvo22}
the evolution of HMFs does not depend significantly on the turbulent
contribution to $\alpha_{\mathrm{eff}}$. Nevertheless, we use the
correct $\alpha$-dynamo parameter here.

The analog of the CME for HMFs is accounted for in $\alpha_{\mathrm{eff}}$
in Eq.~(\ref{eq:etaalphaY}). We found in Ref.~\refcite{DvoSem13}
that its contribution to the $\alpha$-dynamo parameter depends on
the asymmetries of right and left leptons,
\begin{equation}\label{eq:alpha}
  \alpha_{\mathrm{Y}}(\tilde{\eta})=\frac{\alpha'}{\pi\sigma_{c}}
  \left[
    \xi_{e\mathrm{R}}(\tilde{\eta})-\frac{\xi_{e\mathrm{L}}(\tilde{\eta})}{2}
  \right].
\end{equation}
We use the correct sign in the left asymmetry term in Eq.~(\ref{eq:alpha}). The most general expression for the $\alpha$-dynamo parameter in Eq.~\eqref{eq:alpha} was found in Ref.~\refcite{KamLon16} to contain the quarks contribution, $\sim\alpha'\times\text{BAU}/\pi\sigma_c$, in the right hand side, where the expression for BAU is given in Eq.~\eqref{eq:BAUgen} below. In our analysis, we neglect this term since it is smaller compared to the contribution of left and right electrons.

\subsection{Seed spectrum of HMFs and initial asymmetries\label{subsec:SEEDSPEC}}

The numerical analysis of Eq.~(\ref{eq:HMFsys}) requires the initial
condition for HMFs. In Refs.~\refcite{DvoSem21,Dvo22}, we took the
Kolmogorov seed spectrum at $T_{\mathrm{RL}}$. Thus, we had to impose
the minimal momentum $\tilde{k}_{\mathrm{min}}$ to avoid the singularity
of the Kolmogorov spectrum at $\tilde{k}=0$. The value of $\tilde{k}_{\mathrm{min}}$,
which is associated to the maximal length scale, was taken to be the
reciprocal horizon size at $T_{\mathrm{RL}}$: $\tilde{k}_{\mathrm{min}}\sim T_{\mathrm{RL}}/\tilde{M}_{\mathrm{Pl}}\approx10^{-14}$.

However, as mentioned in Ref.~\refcite{Bra17}, the spectrum of HMFs
can be vanishing at distances greater than the horizon size. It leads
to the composite seed spectrum,
\begin{equation}\label{eq:seedspecE}
  \tilde{\mathcal{E}}_{{\rm B_{\mathrm{Y}}}}^{(0)}(\tilde{k})\sim
  \begin{cases}
    \tilde{k}^{n_{\mathrm{B}}}, & 0<\tilde{k}<\tilde{k}_{\star},
    \\
    \tilde{k}^{n_{\mathrm{K}}}, & \tilde{k}_{\star}<\tilde{k}<\tilde{k}_{\mathrm{max}},
  \end{cases}
\end{equation}
i.e. $\tilde{\mathcal{E}}_{{\rm B_{\mathrm{Y}}}}^{(0)}(\tilde{k})$
is of the Batchelor type with $n_{\mathrm{B}}=4$ at small momenta,
and $\tilde{\mathcal{E}}_{{\rm B_{\mathrm{Y}}}}^{(0)}(\tilde{k})$
is of the Kolmogorov type with $n_{\mathrm{K}}=-5/3$ at great momenta.
The border momentum $\tilde{k}_{\star}$ was found in Ref.~\refcite{Bra17}
to be related to the reciprocal horizon size defined above: $\tilde{k}_{\star}=T_{\mathrm{RL}}/\gamma_{\star}\tilde{M}_{\mathrm{Pl}}$.
We shall vary the parameter $\gamma_{\star}$ in the range $10^{-2}<\gamma_{\star}<10^{-3}$.
The seed spectrum of the magnetic helicity can be taken in the form,
$\mathcal{\tilde{H}}_{{\rm B_{\mathrm{Y}}}}^{(0)}(\tilde{k})=2q\mathcal{\tilde{E}}_{{\rm B_{\mathrm{Y}}}}^{(0)}(\tilde{k})/\tilde{k}$,
where $0\leq q\leq 1$ is the phenomenological parameter fixing the helicity
of a seed HMF.

The normalization constant in the seed spectrum can be found using
Eq.~(\ref{eq:seedspecE}) and the new variables in Eq.~(\ref{eq:newvar}),
\begin{equation}
  \frac{\tilde{B}_{0}^{2}}{2}=
  \int_{0}^{\tilde{k}_{\mathrm{max}}}\tilde{\mathcal{E}}_{{\rm B_{\mathrm{Y}}}}^{(0)}(\tilde{k})\mathrm{d}\tilde{k}=
  \frac{\pi^{2}\tilde{k}_{\mathrm{max}}^{2}}{6\alpha^{\prime2}}
  \left[
    C_{\mathrm{B}}\int_{0}^{\kappa_{\star}}\kappa^{n_{\mathrm{B}}}\mathrm{d}\kappa+
    C_{\mathrm{K}}\int_{\kappa_{\star}}^{1}\kappa^{n_{\mathrm{K}}}\mathrm{d}\kappa
  \right],
\end{equation}
where $\tilde{B}_{0}$ is the conformal seed HMF. Moreover,
we request that the seed spectrum is continuous, i.e. $C_{\mathrm{B}}\kappa_{\star}^{n_{\mathrm{B}}}=C_{\mathrm{K}}\kappa_{\star}^{n_{\mathrm{K}}}$.
These conditions define constants $C_{\mathrm{B,K}}$. Finally, we
get that the seed spectrum in the new variables in Eq.~(\ref{eq:newvar}) reads
\begin{equation}\label{eq:seedR0}
  R_{0}(\kappa)=\frac{3\alpha^{\prime2}\tilde{B}_{0}^{2}
  (1+n_{\mathrm{K}})}{\pi^{2}\tilde{k}_{\mathrm{max}}^{2}}
  \left(
    \frac{n_{\mathrm{K}}-n_{\mathrm{B}}}{1+n_{\mathrm{B}}}+\frac{1}{\kappa_{\star}^{1+n_\mathrm{K}}}
  \right)^{-1}
  \times
  \begin{cases}
    \frac{\kappa^{n_{\mathrm{B}}}}{\kappa_{\star}^{1+n_{\mathrm{B}}}}, & 0<\kappa<\kappa_{\star},
    \\
    \frac{\kappa^{n_{\mathrm{K}}}}{\kappa_{\star}^{1+n_{\mathrm{K}}}}, & \kappa_{\star}<\kappa<1,
  \end{cases}
\end{equation}
where $\kappa_{\star}=\tilde{k}_{\star}/\tilde{k}_{\mathrm{max}}$.
Obviously that $0<\kappa_{\star}<1$. The new helicity spectrum corresponding to
Eq.~(\ref{eq:newvar}) is $H_{0}(\kappa)=qR_{0}(\kappa)/\kappa$.

Besides the seed spectra, we should set the initial asymmetries in
Eq.~(\ref{eq:HMFsys}). We take that $\xi_{e\mathrm{L}}=\xi_{0}=0$
and $\xi_{e\mathrm{R}}=10^{-10}$. It means that, if we study the
BAU production in the system (see Sec.~\ref{sec:BAU} below),
the main contribution to BAU stems from the right electrons component. Analogous initial
asymmetries were considered in Refs.~\refcite{DvoSem21,Dvo22}.

\subsection{Numerical simulations of the HMF behavior\label{subsec:NUMHMF}}

In this section, we present the numerical solution of Eq.~(\ref{eq:HMFsys})
with the initial condition formulated in Sec.~\ref{subsec:SEEDSPEC}.
For this purpose, we rewrite the system in the form in Eq.~(\ref{eq:newsys}).
We take that seed HMFs are maximally helical, i.e. $q=1$. Indeed,
we have shown in Refs.~\refcite{DvoSem21,Dvo22} that the behavior of
HMFs only slightly depends on $q$.

\begin{figure}
  \centering
  \subfigure[]
  {\label{fig:1a}
  \includegraphics[scale=.3]{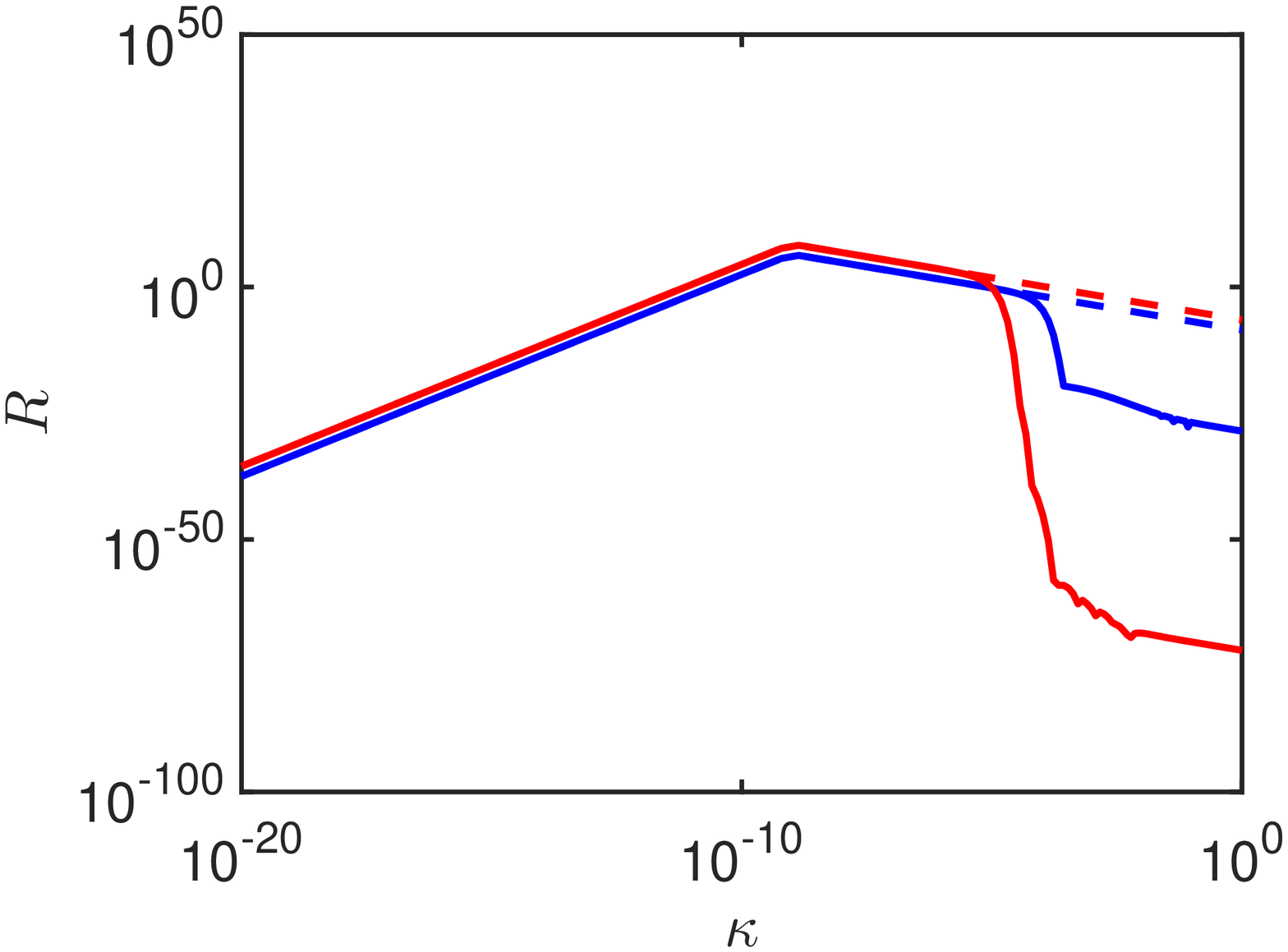}}
  \hskip-.5cm
  \subfigure[]
  {\label{fig:1b}
  \includegraphics[scale=.3]{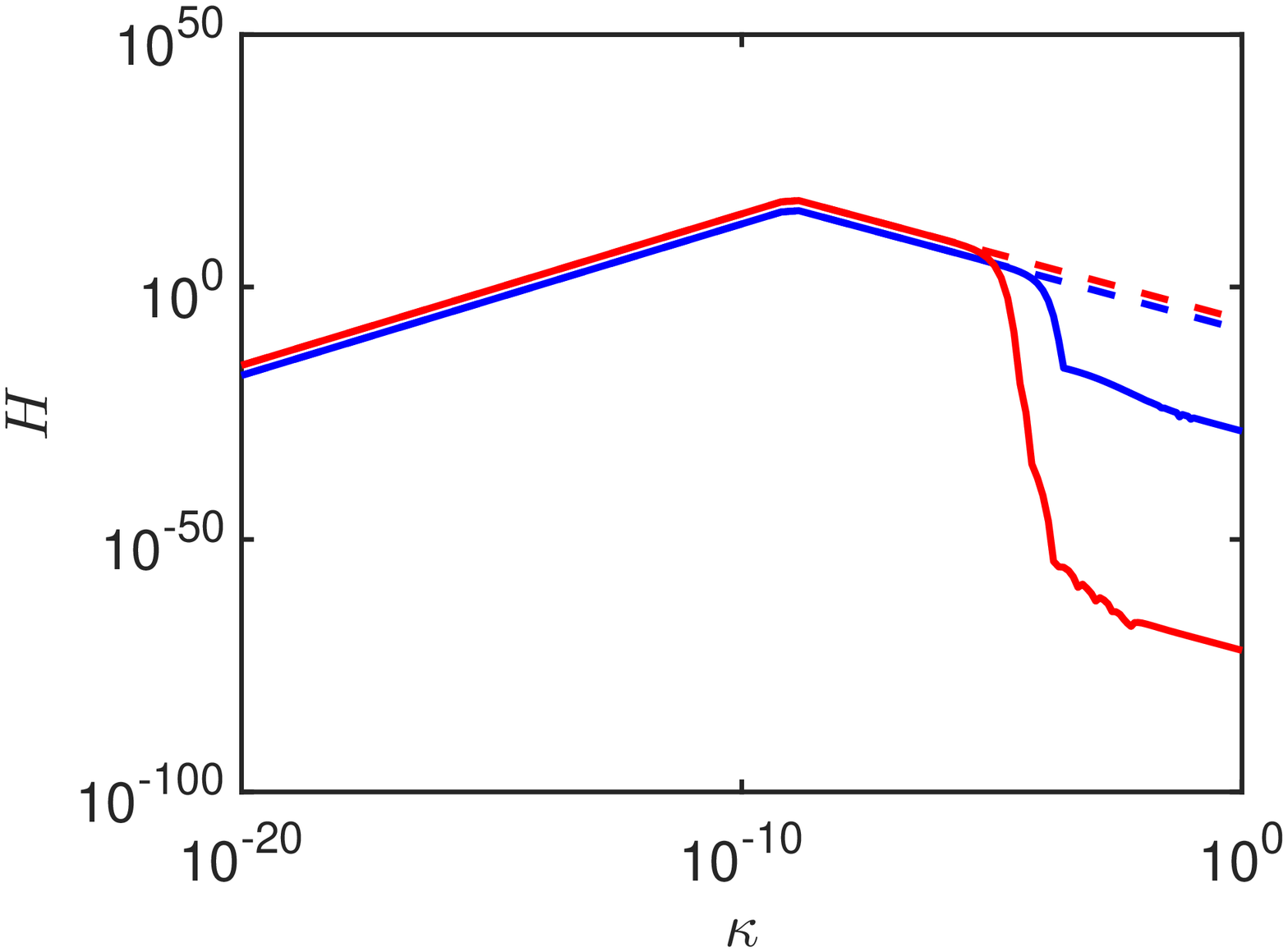}}
  \\
  \vskip-.3cm
  \subfigure[]
  {\label{fig:1c}
  \includegraphics[scale=.3]{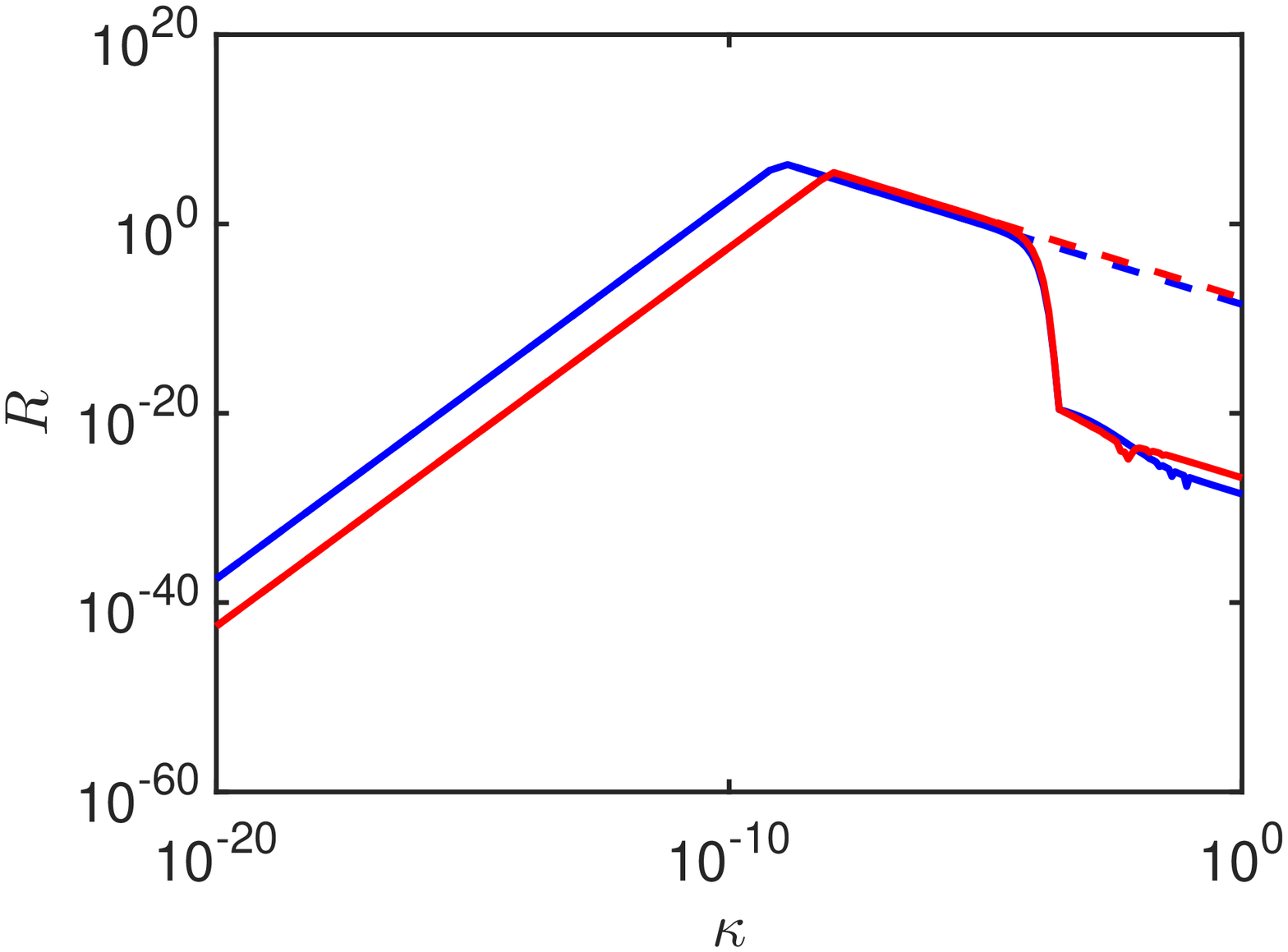}}
  \hskip-.5cm
  \subfigure[]
  {\label{fig:1d}
  \includegraphics[scale=.3]{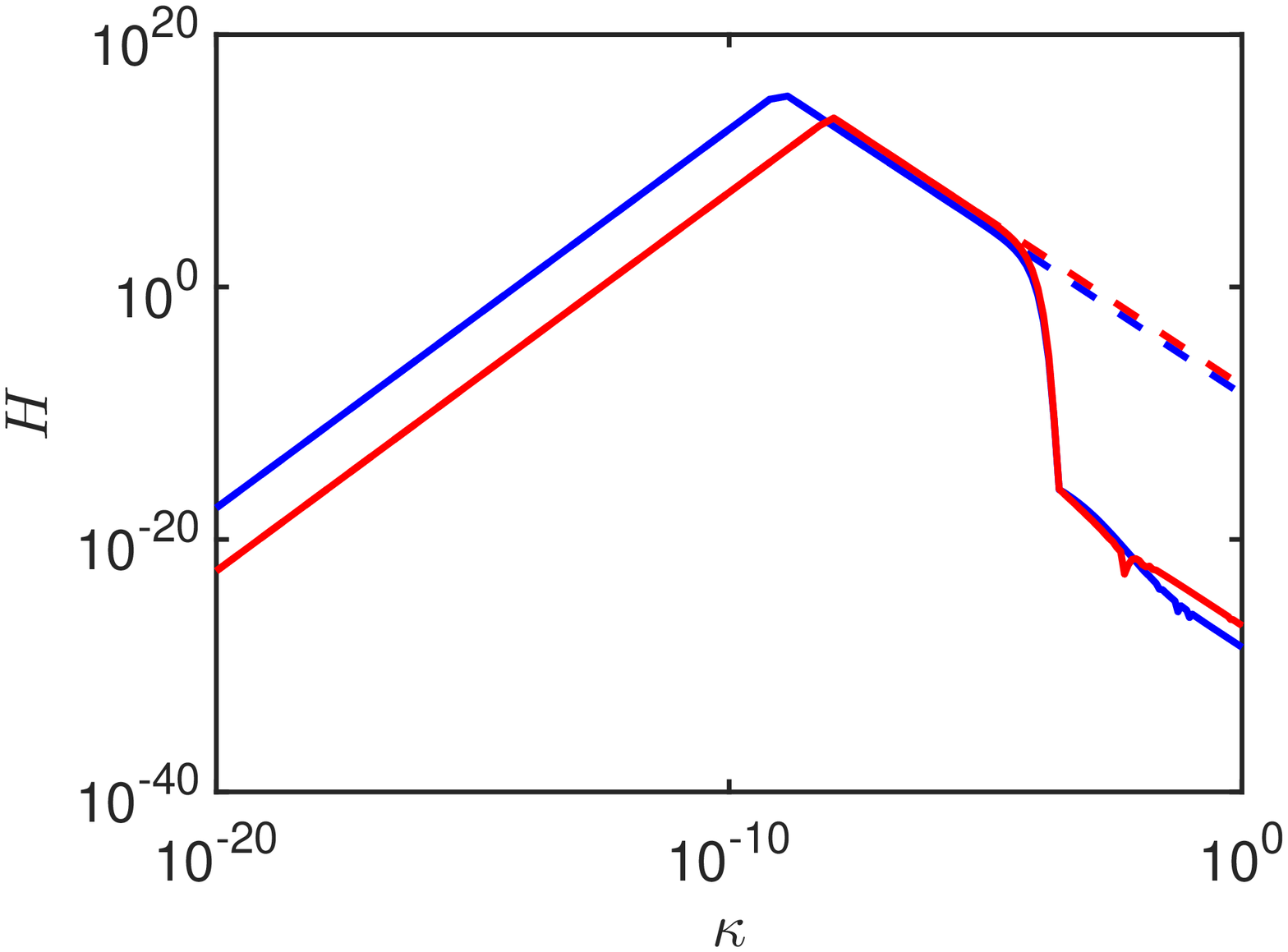}}
  \\
  \vskip-.3cm
  \subfigure[]
  {\label{fig:1e}
  \includegraphics[scale=.3]{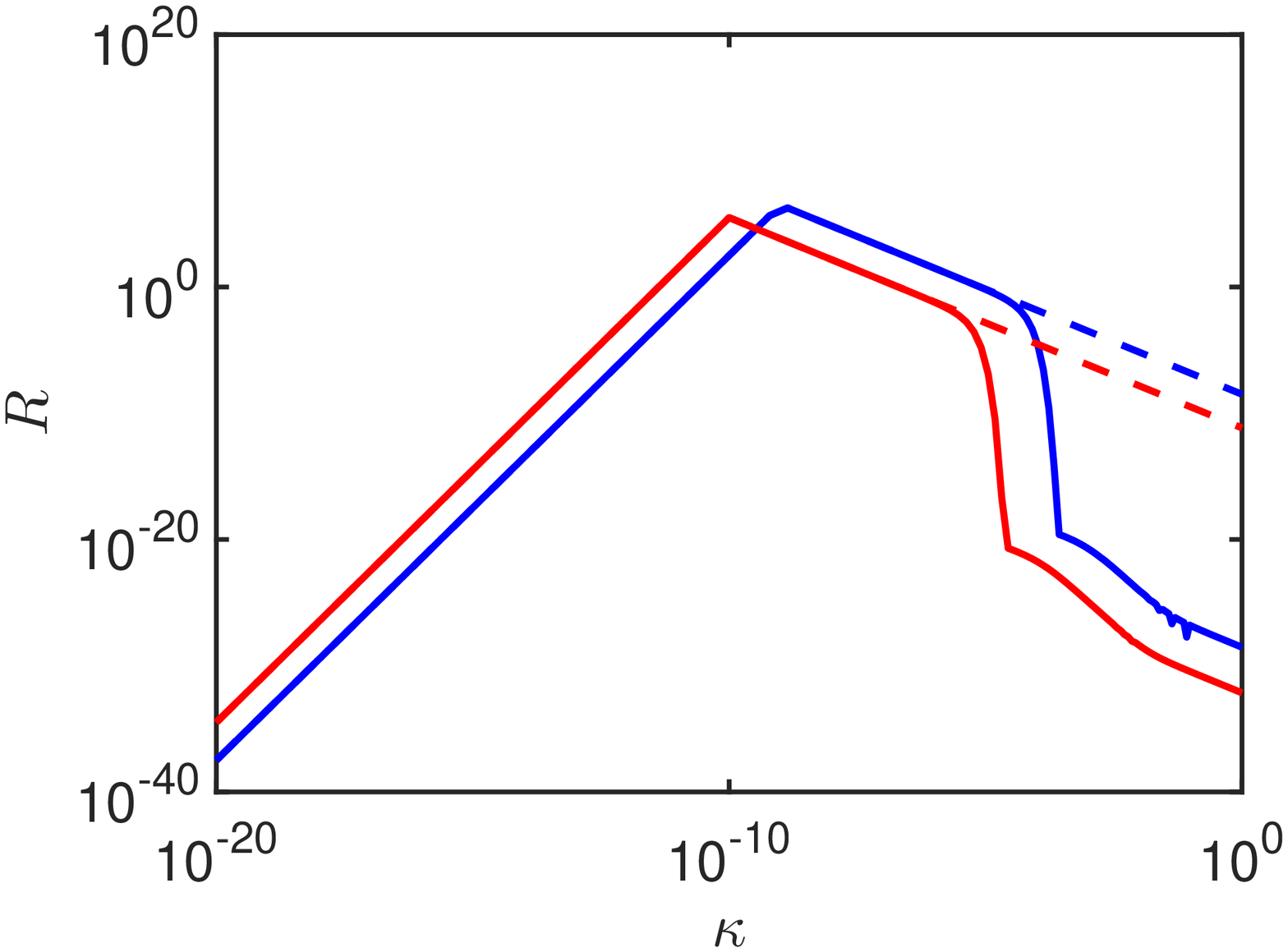}}
  \hskip-.5cm
  \subfigure[]
  {\label{fig:1f}
  \includegraphics[scale=.3]{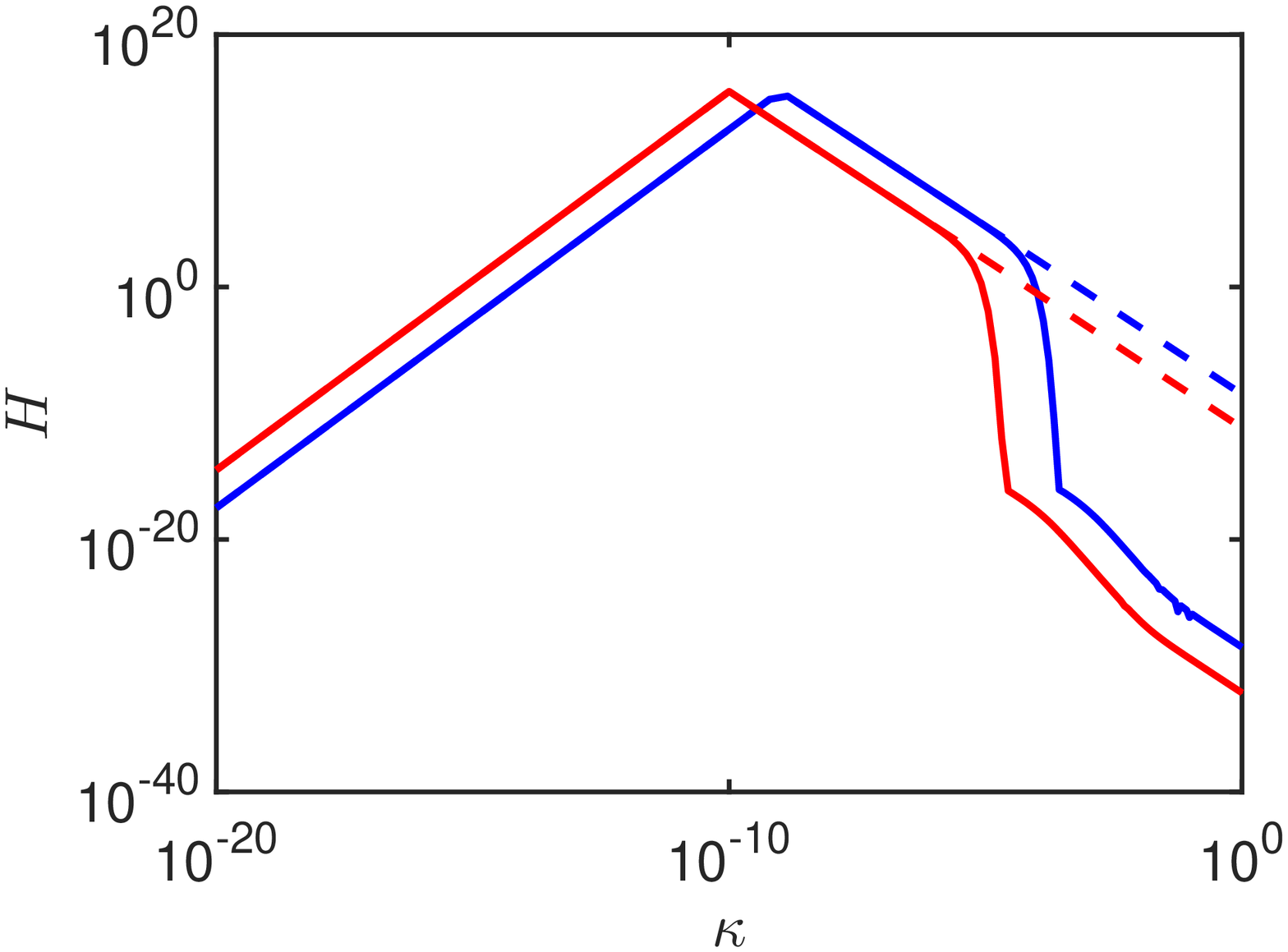}}
  \protect
\caption{The spectra versus $\kappa$ for different parameters of the system.
Solid lines are the numerical solution of Eq.~(\ref{eq:newsys})
at EWPT. Dashed lines are the seed spectra at $T_{\mathrm{RL}}$ corresponding
to Eq.~(\ref{eq:seedR0}). Panels (a), (c), and (e): the spectra
of the hypermagnetic energy density $R(\kappa)$ in Eq.~(\ref{eq:newvar}).
Panels (b), (d), and (f): the spectra of the hypermagnetic helicity
density $H(\kappa)$ in Eq.~(\ref{eq:newvar}). Blue lines in panels
(a) and (b) correspond to $\tilde{B}_{\mathrm{Y}}^{(0)}=1.4\times10^{-2}$
and red ones to $\tilde{B}_{\mathrm{Y}}^{(0)}=1.4\times10^{-1}$.
Additionally $\gamma_{\star}=10^{-2}$ and $\tilde{k}_{\mathrm{max}}=10^{-3}$
in panels (a) and (b). Blue lines in panels (c) and (d) correspond
to $\gamma_{\star}=10^{-2}$ and red ones to $\gamma_{\star}=10^{-3}$.
Additionally $\tilde{B}_{\mathrm{Y}}^{(0)}=1.4\times10^{-2}$ and
$\tilde{k}_{\mathrm{max}}=10^{-3}$ in panels (c) and (d). Blue lines
in panels (e) and (f) correspond to $\tilde{k}_{\mathrm{max}}=10^{-3}$
and red ones to $\tilde{k}_{\mathrm{max}}=10^{-2}$. Additionally
$\tilde{B}_{\mathrm{Y}}^{(0)}=1.4\times10^{-2}$ and $\gamma_{\star}=10^{-2}$
in panels (e) and (f).\label{fig:spectra}}
\end{figure}

We demonstrate in Fig.~\ref{fig:spectra} the spectra of the energy
density $R\propto\tilde{\mathcal{E}}_{{\rm B_{\mathrm{Y}}}}$ and
of the helicity density $H\propto\tilde{\mathcal{H}}_{{\rm B_{\mathrm{Y}}}}$
versus $\kappa$ for different parameters of the system. We show both
the seed spectra and the spectra at EWPT in Fig.~\ref{fig:spectra}.
In particular, we fix $\gamma_{\star}$ and $\tilde{k}_{\mathrm{max}}$,
and change $\tilde{B}_{\mathrm{Y}}^{(0)}$ in Figs.~\ref{fig:1a}
and~\ref{fig:1b}; fix $\tilde{B}_{\mathrm{Y}}^{(0)}$ and
$\tilde{k}_{\mathrm{max}}$, and change $\gamma_{\star}$ in Figs.~\ref{fig:1c}
and~\ref{fig:1d}; and, finally, fix $\tilde{B}_{\mathrm{Y}}^{(0)}$
and $\gamma_{\star}$, and change $\tilde{k}_{\mathrm{max}}$ in Figs.~\ref{fig:1e}
and~\ref{fig:1f}. Irregular parts of lines at the very great
momenta $\kappa\lesssim1$ are because of the inexactutude
of numerical simulations.

We can see in Fig.~\ref{fig:spectra} that the small momenta tails
of the spectra are unchanged and coincide with the seed Batchelor
spectra, which are also shown Fig.~\ref{fig:spectra} by dashed lines.
The Kolmogorov parts of the spectra, corresponding to great momenta,
are mainly affected in the evolution. This behavior of the spectra
qualitatively resembles that found in Ref.~\refcite{Dvo22}. It results
from the fact that the major contribution to the evolution of HMFs
is from the diffusion terms, $\propto-\kappa^{2}R$ and $\propto-\kappa^{2}H$,
in the right hand side of Eq.~\ref{eq:newsys}. The dynamo amplification
of HMFs is not effective in the present system.

\begin{figure}
  \centering
  \subfigure[]
  {\label{fig:2a}
  \includegraphics[scale=.3]{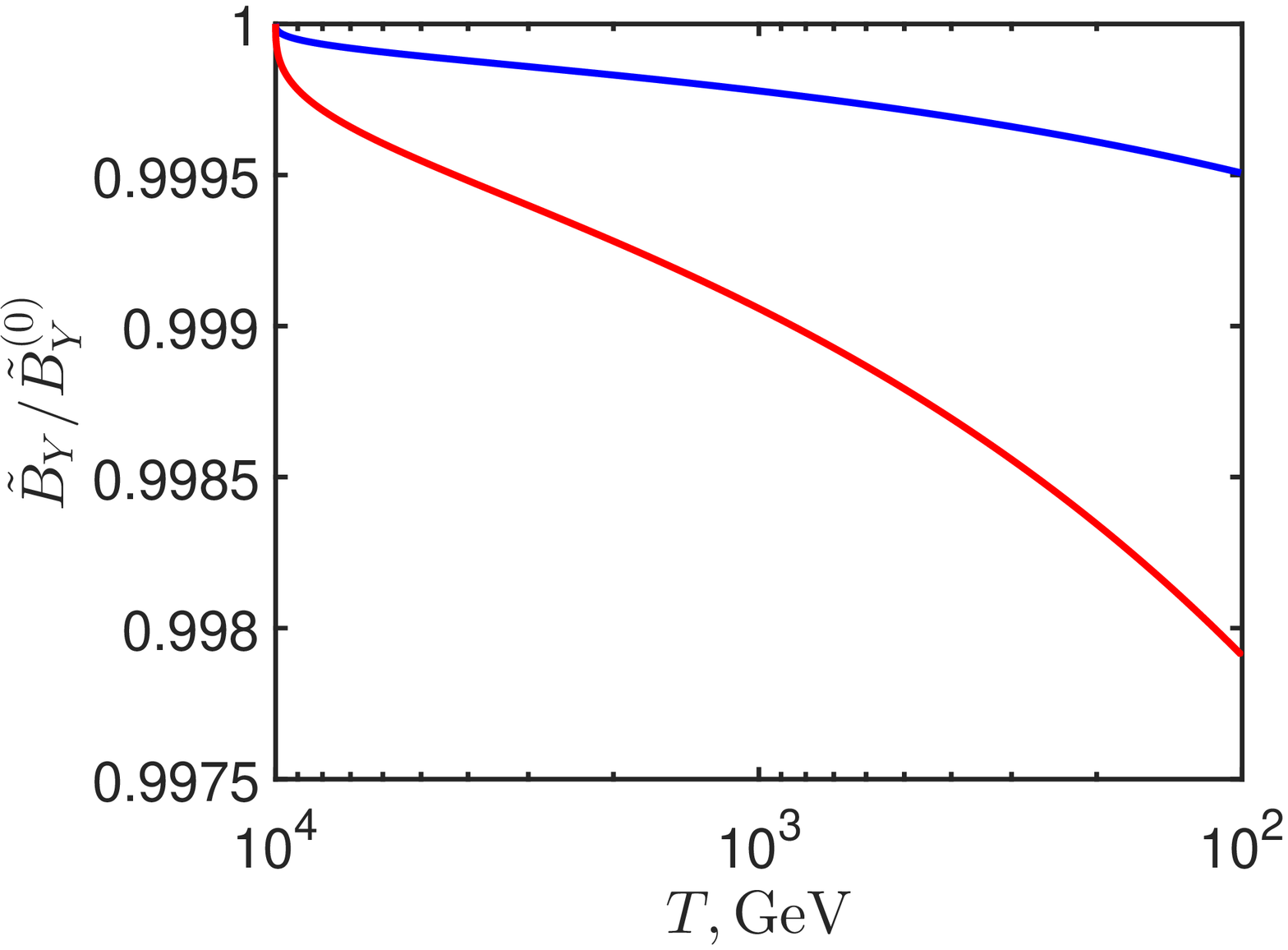}}
  \hskip-.5cm
  \subfigure[]
  {\label{fig:2b}
  \includegraphics[scale=.3]{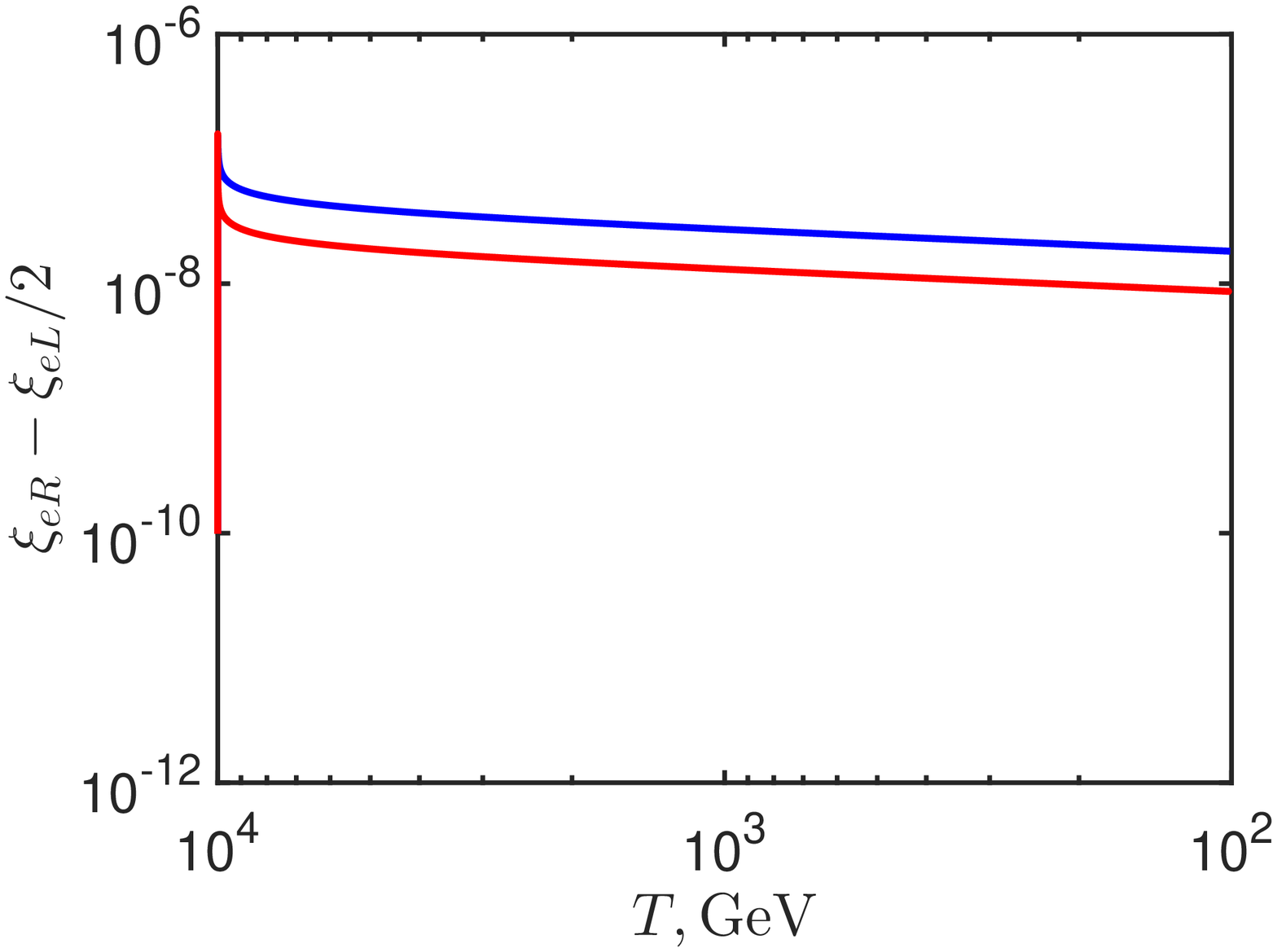}}
  \\
  \vskip-.3cm
  \subfigure[]
  {\label{fig:2c}
  \includegraphics[scale=.3]{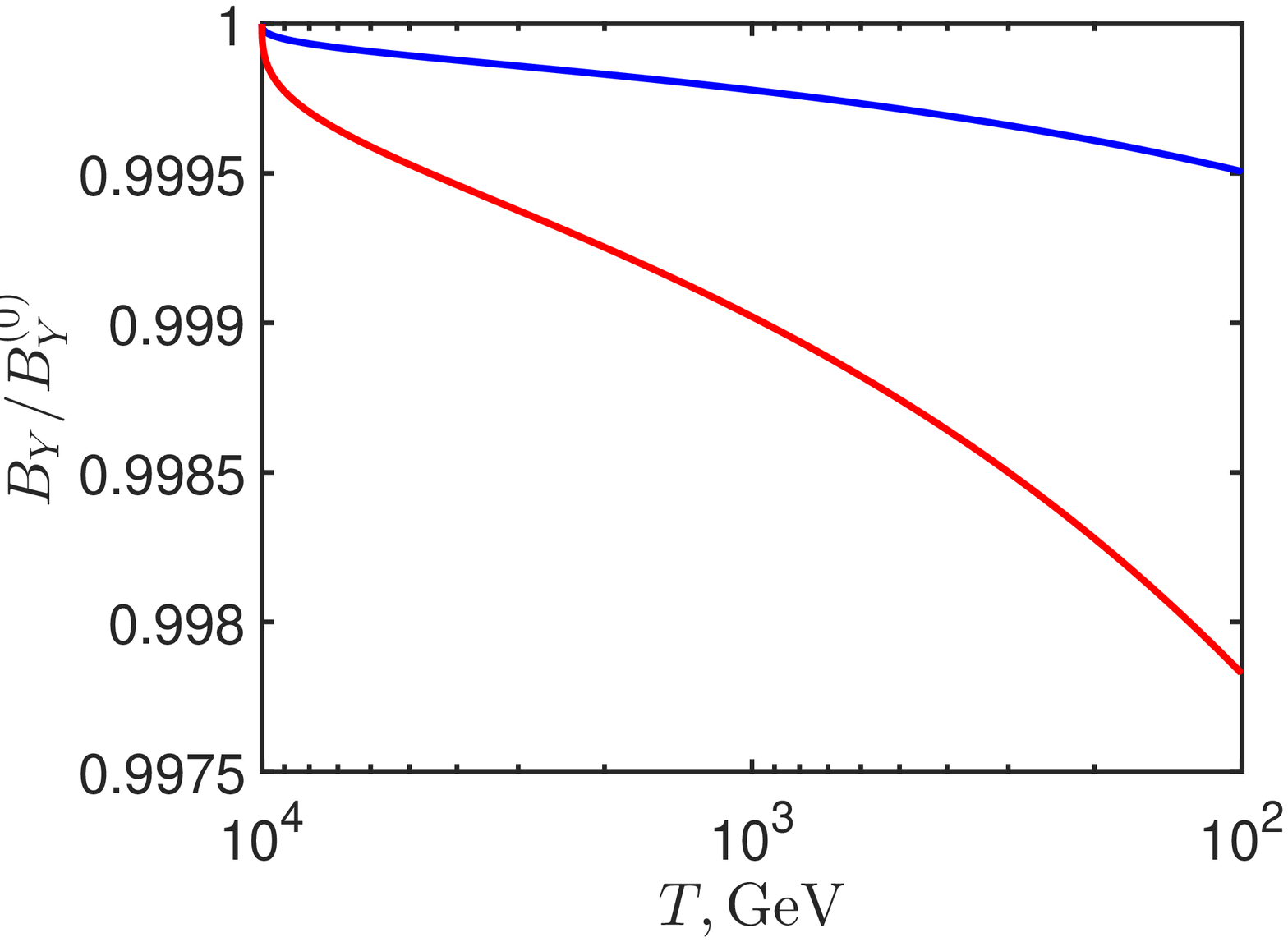}}
  \hskip-.5cm
  \subfigure[]
  {\label{fig:2d}
  \includegraphics[scale=.3]{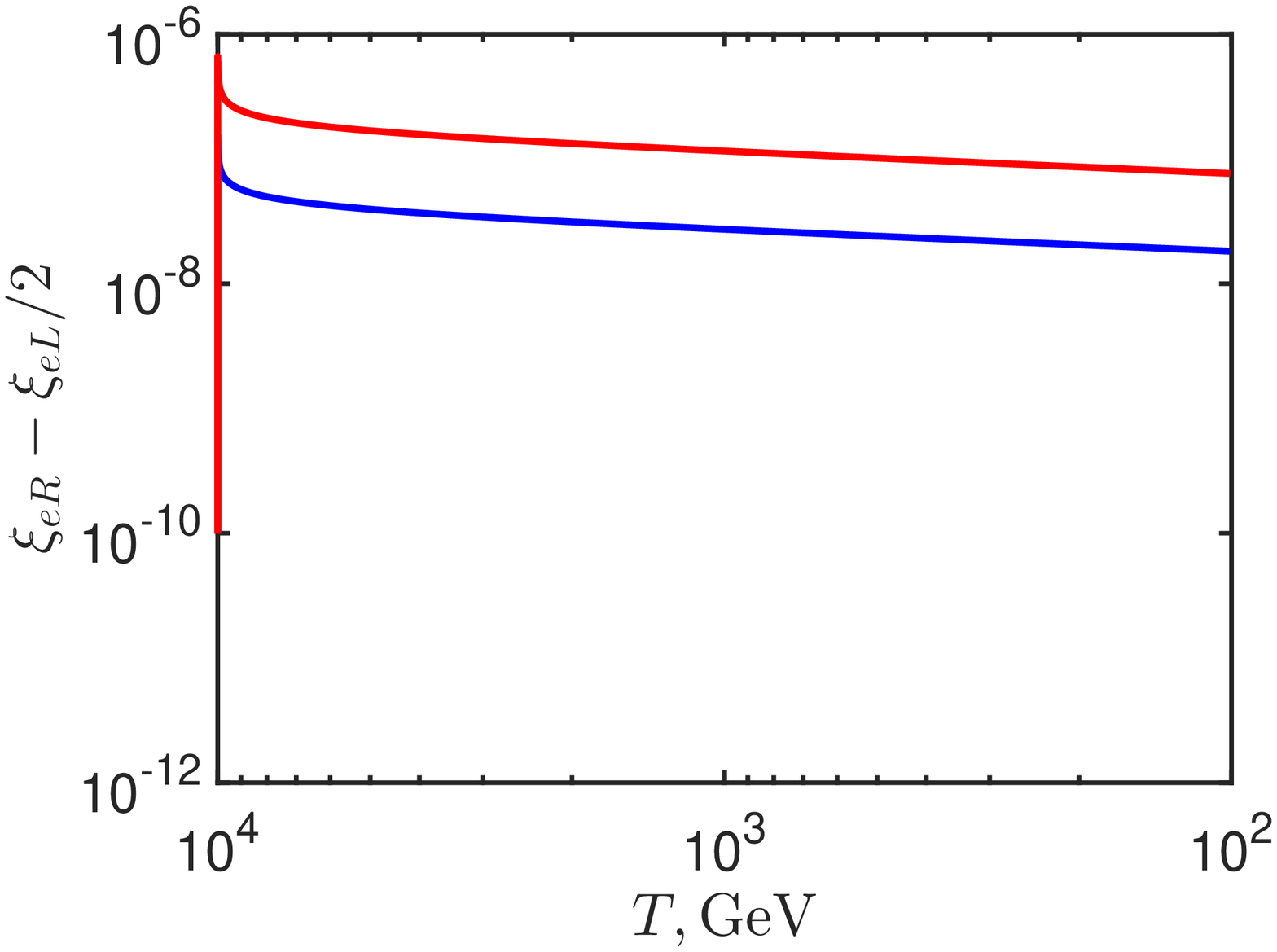}}
  \\
  \vskip-.3cm
  \subfigure[]
  {\label{fig:2e}
  \includegraphics[scale=.3]{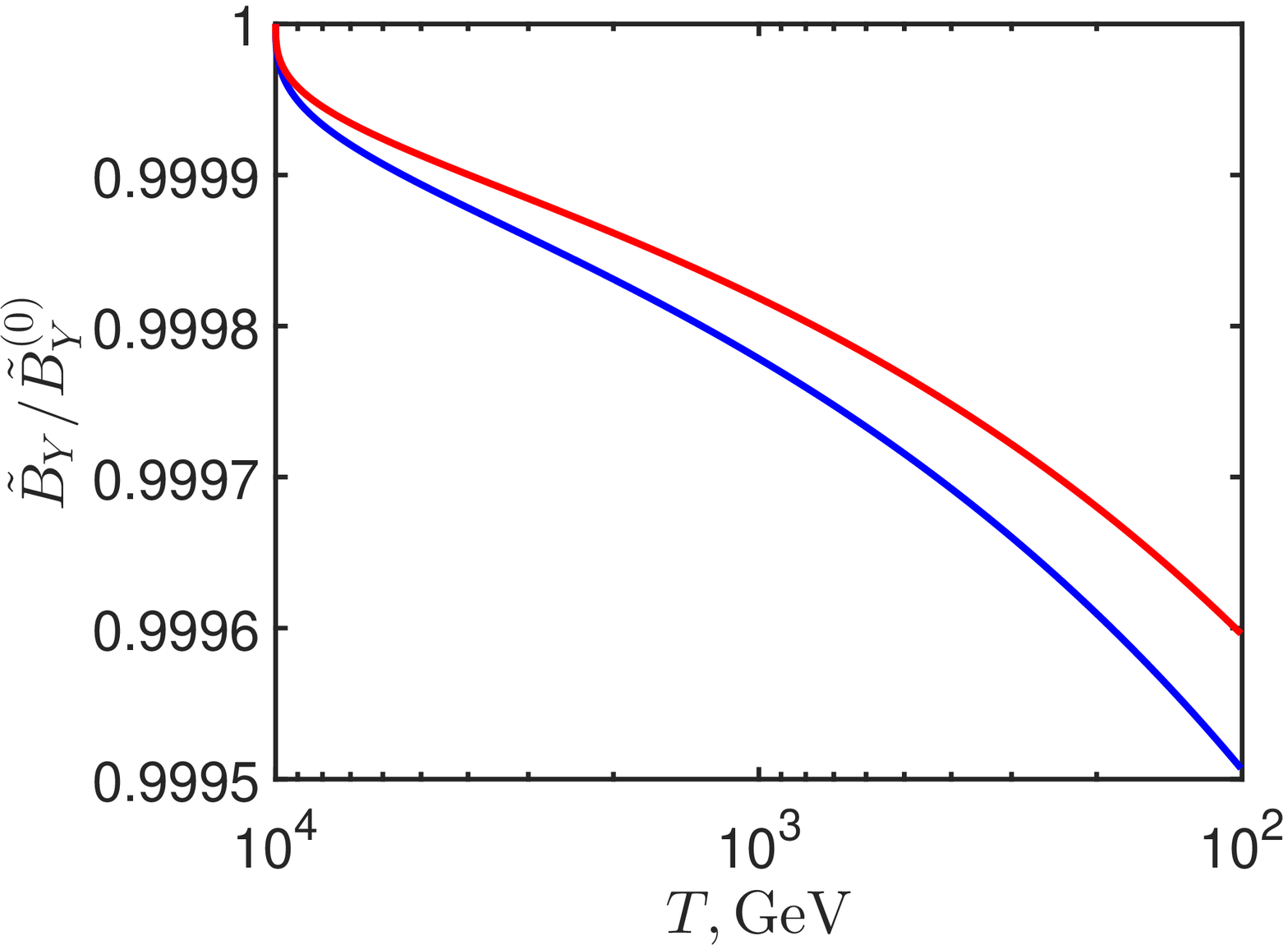}}
  \hskip-.5cm
  \subfigure[]
  {\label{fig:2f}
  \includegraphics[scale=.3]{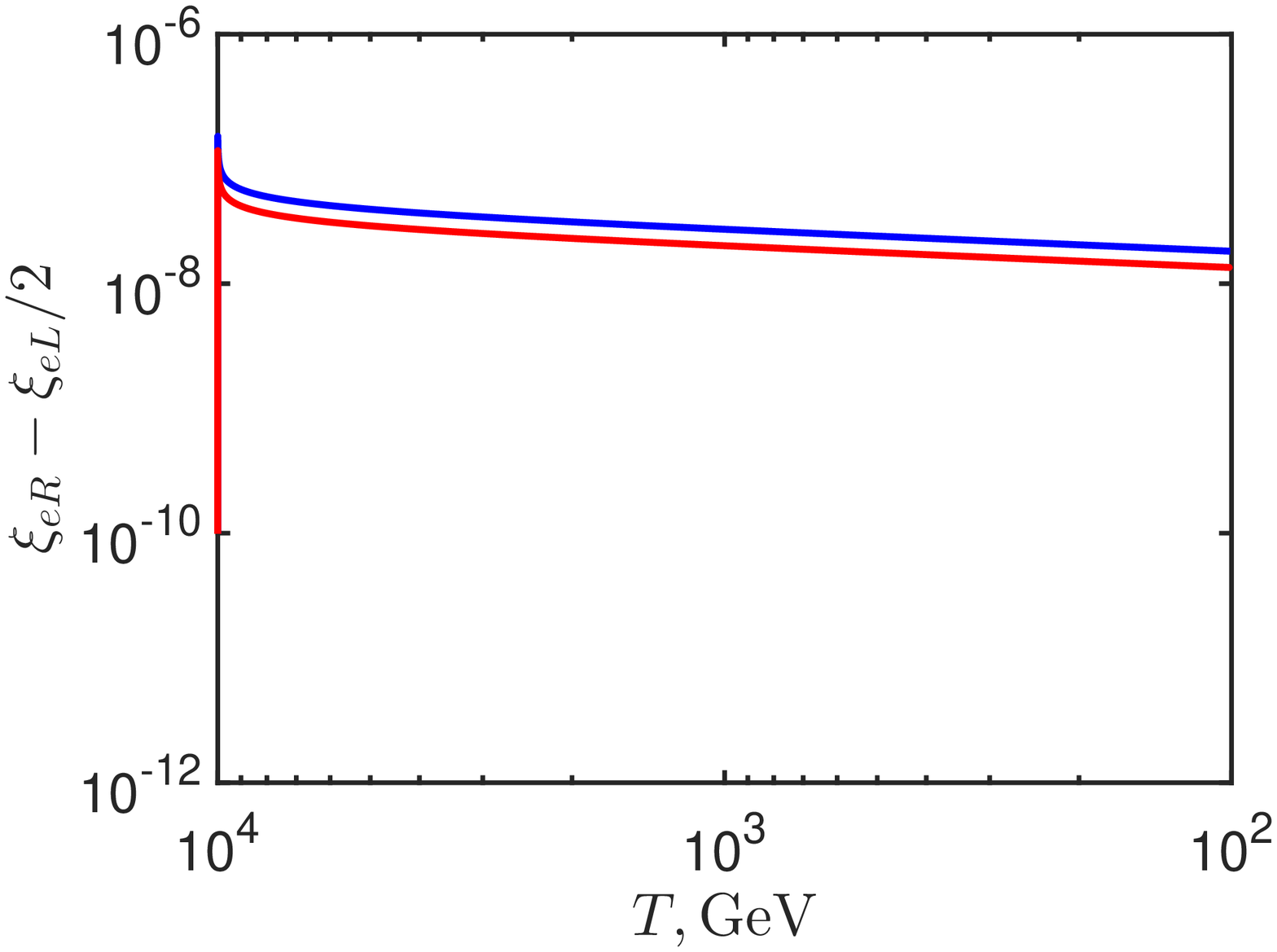}}
  \protect
\caption{The strength of HMF $\tilde{B}_{\mathrm{Y}}$ normalized by the seed
strength $\tilde{B}_{\mathrm{Y}}^{(0)}$ versus the plasma temperature
$T$ is shown in panels (a), (c), and (e) for different parameters
of the system. The chiral $\alpha$-dynamo parameter in Eq.~(\ref{eq:alpha}),
$\alpha_{\mathrm{Y}}\propto\xi_{e\mathrm{R}}-\xi_{e\mathrm{L}}/2$,
depending on the plasma temperature $T$, is depicted in panels (b),
(d), and (f) for different parameters of the system. Blue line in
panels (a) and (b) correspond to $\tilde{B}_{\mathrm{Y}}^{(0)}=1.4\times10^{-2}$
and red one to $\tilde{B}_{\mathrm{Y}}^{(0)}=1.4\times10^{-1}$.
Additionally $\gamma_{\star}=10^{-2}$ and $\tilde{k}_{\mathrm{max}}=10^{-3}$
in panels (a) and (b). Blue line in panels (c) and (d) correspond
to $\gamma_{\star}=10^{-2}$ and red one to $\gamma_{\star}=10^{-3}$.
Additionally $\tilde{B}_{\mathrm{Y}}^{(0)}=1.4\times10^{-2}$ and
$\tilde{k}_{\mathrm{max}}=10^{-3}$ in panels (c) and (d). Blue line
in panels (e) and (f) correspond to $\tilde{k}_{\mathrm{max}}=10^{-3}$
and red one to $\tilde{k}_{\mathrm{max}}=10^{-2}$. Additionally
$\tilde{B}_{\mathrm{Y}}^{(0)}=1.4\times10^{-2}$ and $\gamma_{\star}=10^{-2}$
in panels (e) and (f).\label{fig:B}}
\end{figure}

We show the evolution of the HMFs strength $\tilde{B}_{\mathrm{Y}}$
in Figs.~\ref{fig:2a}, \ref{fig:2c}, and~\ref{fig:2e};
as well as $\alpha_{\mathrm{Y}}\propto\xi_{e\mathrm{R}}-\xi_{e\mathrm{L}}/2$
in Figs.~\ref{fig:2b}, \ref{fig:2d}, and~\ref{fig:2f}.
The values of the parameters of the system in Fig.~\ref{fig:B} are
the same as in the corresponding panels in Fig.~\ref{fig:spectra}.
Qualitatively, the evolution of $\tilde{B}_{\mathrm{Y}}$ and the
asymmetries is similar to that described in Refs.~\refcite{DvoSem21,Dvo22}.

In particular, we can see in Figs.~\ref{fig:2a}, \ref{fig:2c},
and~\ref{fig:2e} that the HMF strength becomes smaller in the
cooling universe. It confirms our claim above that the dynamo amplification
is not effective in this system. The $\alpha$-dynamo parameter Figs.~\ref{fig:2b},
\ref{fig:2d}, and~\ref{fig:2f} has a sharp peak at $T\lesssim T_{\mathrm{RL}}$.
This peak appears mainly owing to $\xi_{e\mathrm{R}}$. It results
from the term $\propto I_{\mathrm{H}}$ in the right hand side of
the equation for $M_{\mathrm{R}}$ in Eq.~(\ref{eq:newsys}).

\section{Production of primordial GWs\label{sec:PRODGWs}}

In this section, we study the generation of relic GWs by HMFs in the
universe cooling down to EWPT. The main formalism for this problem
was developed in Ref.~\refcite{Dvo22}. Here, we just remind the main
steps of that study.

GWs are produced by random HMFs when the energy-momentum tensor of HMFs is
accounted for in the right hand side of the Einstein equation which
is written down in the expanding universe with the Friedmann--Robertson--Walker (FRW)
metric. Using the transverse-traceless gauge and appropriately averaging
the expression for the spectrum of the energy density of GWs $\rho_{\mathrm{GW}}^{(c)}(k,\eta)$,
one gets that it has form~\cite{Dvo22},
\begin{align}\label{eq:rhoGWisotr}
  \rho_{\mathrm{GW}}^{(c)}(k,\eta)= & \frac{t_{\text{Univ}}^{2}G}{4k^{3}\pi^{2}}\eta
  \int_{0}^{\eta}\frac{\mathrm{d}\xi}{(\eta_{0}+\xi)^{2}}
  \int_{0}^{\infty}\frac{\mathrm{d}q}{q^{3}}\int_{|k-q|}^{k+q}\frac{\mathrm{d}p}{p^{3}}
  \nonumber
  \\
  & \times
  \big\{
    [4k^{2}q^{2}+(k^{2}+q^{2}-p^{2})^{2}][4k^{2}p^{2}+(k^{2}-q^{2}+p^{2})^{2}]
    \rho_{\mathrm{Y}}^{(c)}(q,\xi)\rho_{\mathrm{Y}}^{(c)}(p,\xi)
    \nonumber
    \\
    & +
    4k^{2}q^{2}p^{2}(k^{2}+q^{2}-p^{2})(k^{2}-q^{2}+p^{2})
    h_{\mathrm{Y}}^{(c)}(q,\xi)h_{\mathrm{Y}}^{(c)}(p,\xi)
  \big\}.
\end{align}
Here, $\rho_{\mathrm{Y}}^{(c)}(k,\eta)$ and $h_{\mathrm{Y}}^{(c)}(k,\eta)$
are the conformal dimensional spectra of the densities of the HMF
energy and the helicity, which are related to the quantities defined in
Sec.~\ref{sec:EVOLHMF} by $\rho_{\mathrm{Y}}^{(c)}(k,\xi)=\tilde{\mathcal{E}}_{{\rm B_{\mathrm{Y}}}}(\tilde{k},\tilde{\eta})T_{0}^{3}$
and $h_{\mathrm{Y}}^{(c)}(k,\xi)=\tilde{\mathcal{H}}_{{\rm B_{\mathrm{Y}}}}(\tilde{k},\tilde{\eta})T_{0}^{2}$,
where $T_{0}=2.7\,\text{K}$ is the present temperature of the cosmic
microwave background radiation. The dimensional conformal time $\eta$
and the conformal momentum $k$ in Eq.~(\ref{eq:rhoGWisotr}) are $\eta=(2t_{\mathrm{Univ}}T_{0}/\tilde{M}_{\mathrm{Pl}})\tilde{\eta}$
and $k=T_{0}\tilde{k}$, where $t_{\mathrm{Univ}}=1.4\times10^{10}\,\text{yr}$
is the universe age, as well as $\tilde{\eta}$ and $\tilde{k}$ are
defined in Sec.~\ref{sec:EVOLHMF}. In Eq.~(\ref{eq:rhoGWisotr}),
the parameter $\eta_{0}=2t_{\mathrm{Univ}}T_{0}/T_{\mathrm{RL}}$
and $G=M_{\mathrm{Pl}}^{-2}$ is the Newton constant. The total energy
density of GWs is calculated on the basis of Eq.~(\ref{eq:rhoGWisotr}),
\begin{equation}\label{eq:densGW}
  \rho_{\mathrm{GW}}^{(c)}(\eta)=\int_{0}^{\infty}\rho_{\mathrm{GW}}^{(c)}(k,\eta)\mathrm{d}k.
\end{equation}
Note that both $\rho_{\mathrm{GW}}^{(c)}(k,\eta)$ and $\rho_{\mathrm{GW}}^{(c)}(\eta)$
in Eqs.~(\ref{eq:rhoGWisotr}) and~(\ref{eq:densGW}) are conformal.

The total conformal energy density of GWs in Eq.~\eqref{eq:densGW} is obtained from the time component of the effective energy-momentum tensor, $t_{\mu\nu} = \tfrac{1}{32\pi G} \langle \partial_\mu h_{\alpha\beta} \partial_\nu h^{\alpha\beta} \rangle$, as $\rho_{\mathrm{GW}}^{(c)} = a^4 t_{00}$. Here $h_{\alpha\beta}$ is the perturbation of the FRW metric and $a$ is the scale factor.

\subsection{Results for the GW generation and their observability\label{subsec:RESGWs}}

In this section, we analyze the generation of relic GWs driven by
HMFs relying on the results of numerical simulations in Sec.~\ref{subsec:NUMHMF}.
We suppose that $\rho_{\mathrm{GW}}^{(c)}(k,\eta)=0$ at $T_{\mathrm{RL}}$.

\begin{figure}
  \centering
  \subfigure[]
  {\label{fig:3a}
  \includegraphics[scale=.3]{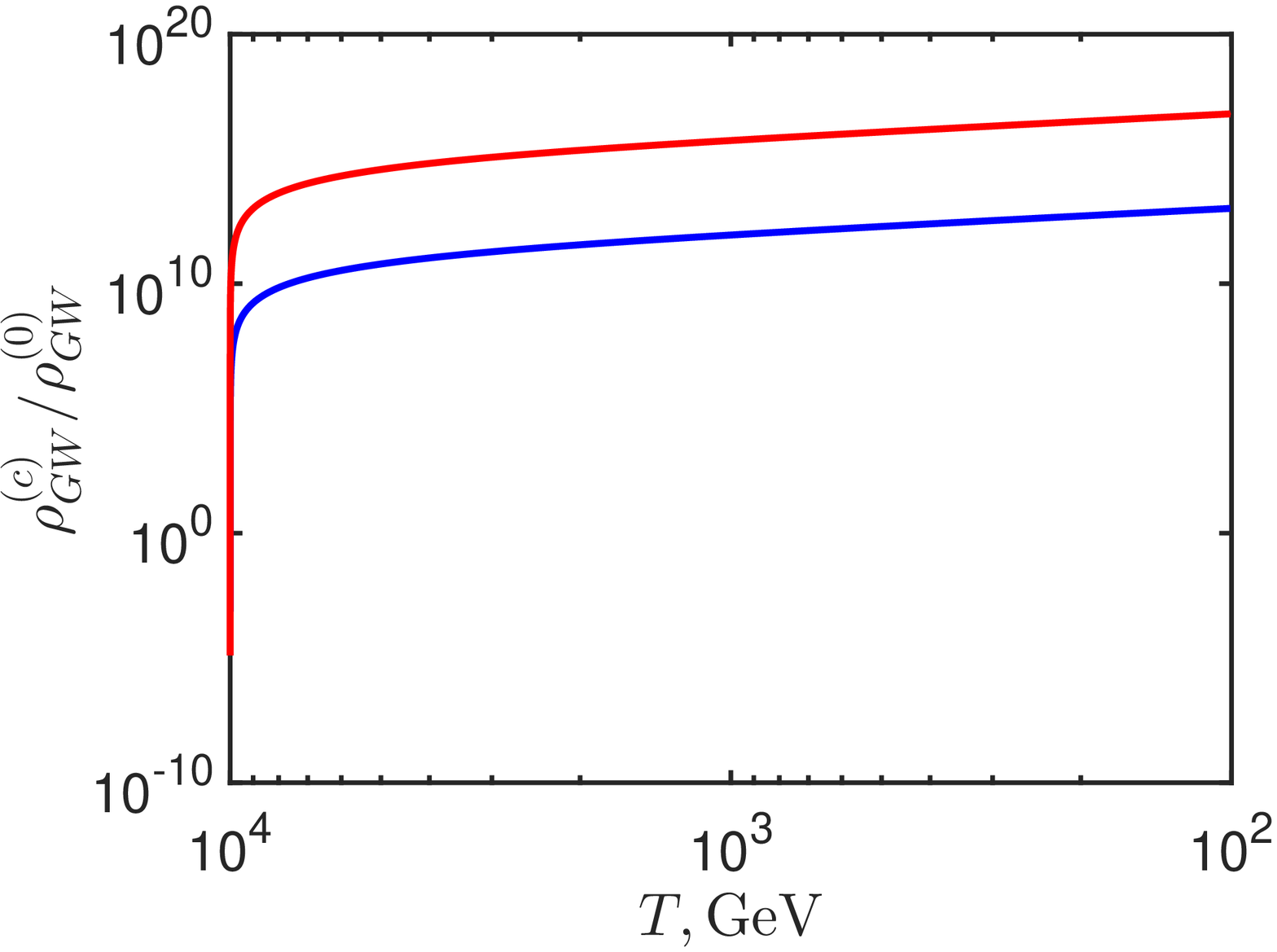}}
  \hskip-.5cm
  \subfigure[]
  {\label{fig:3b}
  \includegraphics[scale=.3]{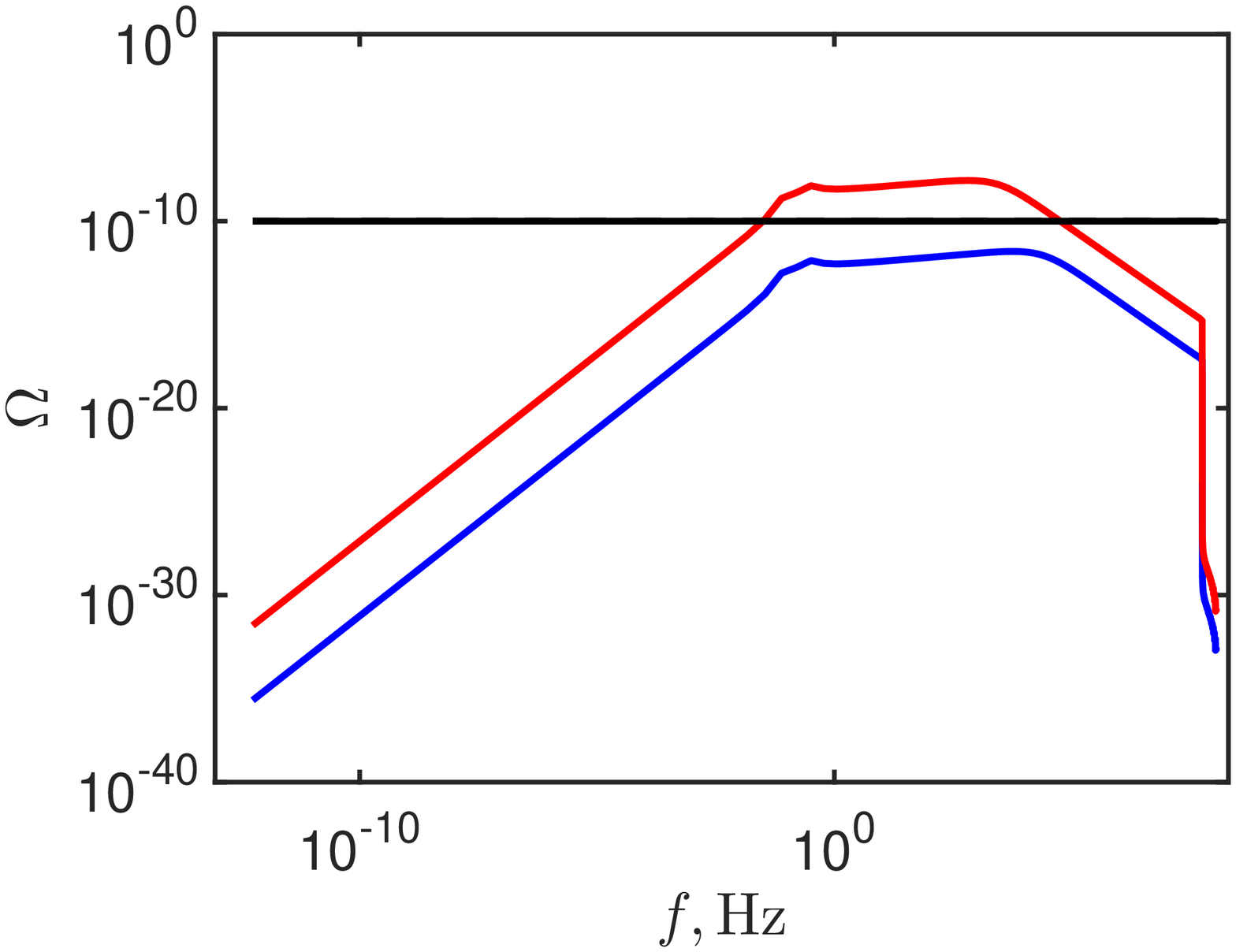}}
  \\
  \vskip-.4cm
  \subfigure[]
  {\label{fig:3c}
  \includegraphics[scale=.3]{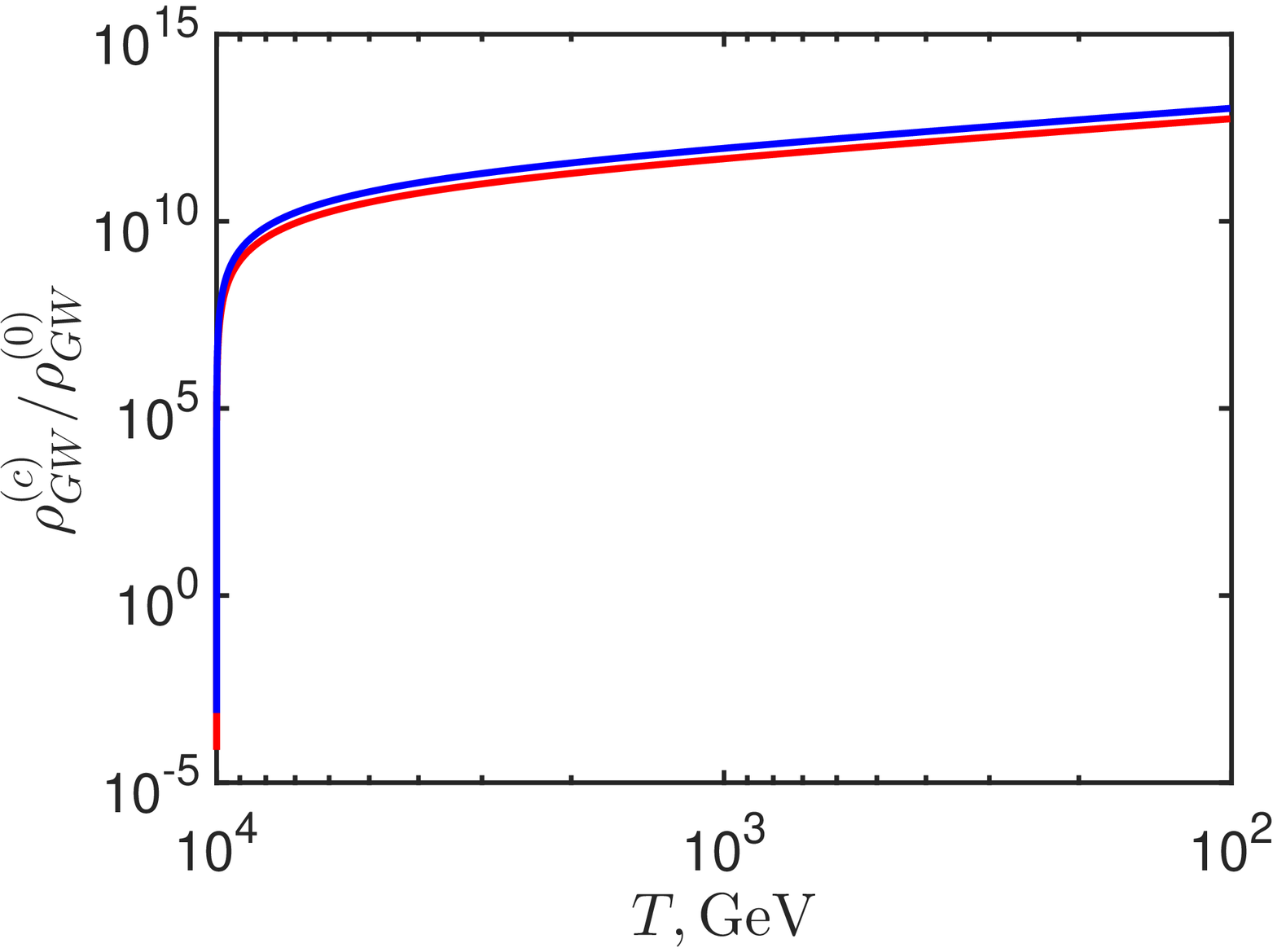}}
  \hskip-.5cm
  \subfigure[]
  {\label{fig:3d}
  \includegraphics[scale=.3]{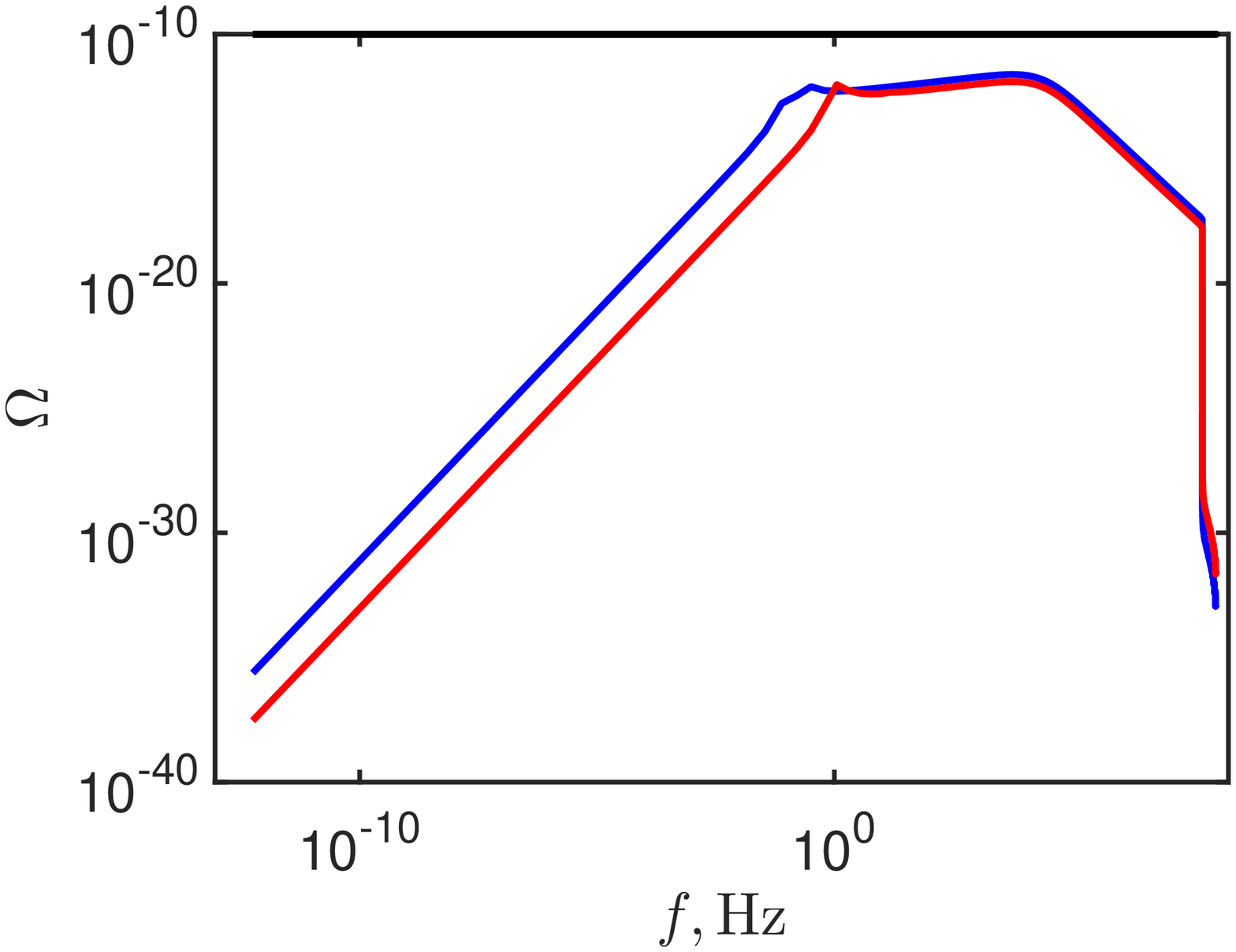}}
  \\
  \vskip-.4cm
  \subfigure[]
  {\label{fig:3e}
  \includegraphics[scale=.3]{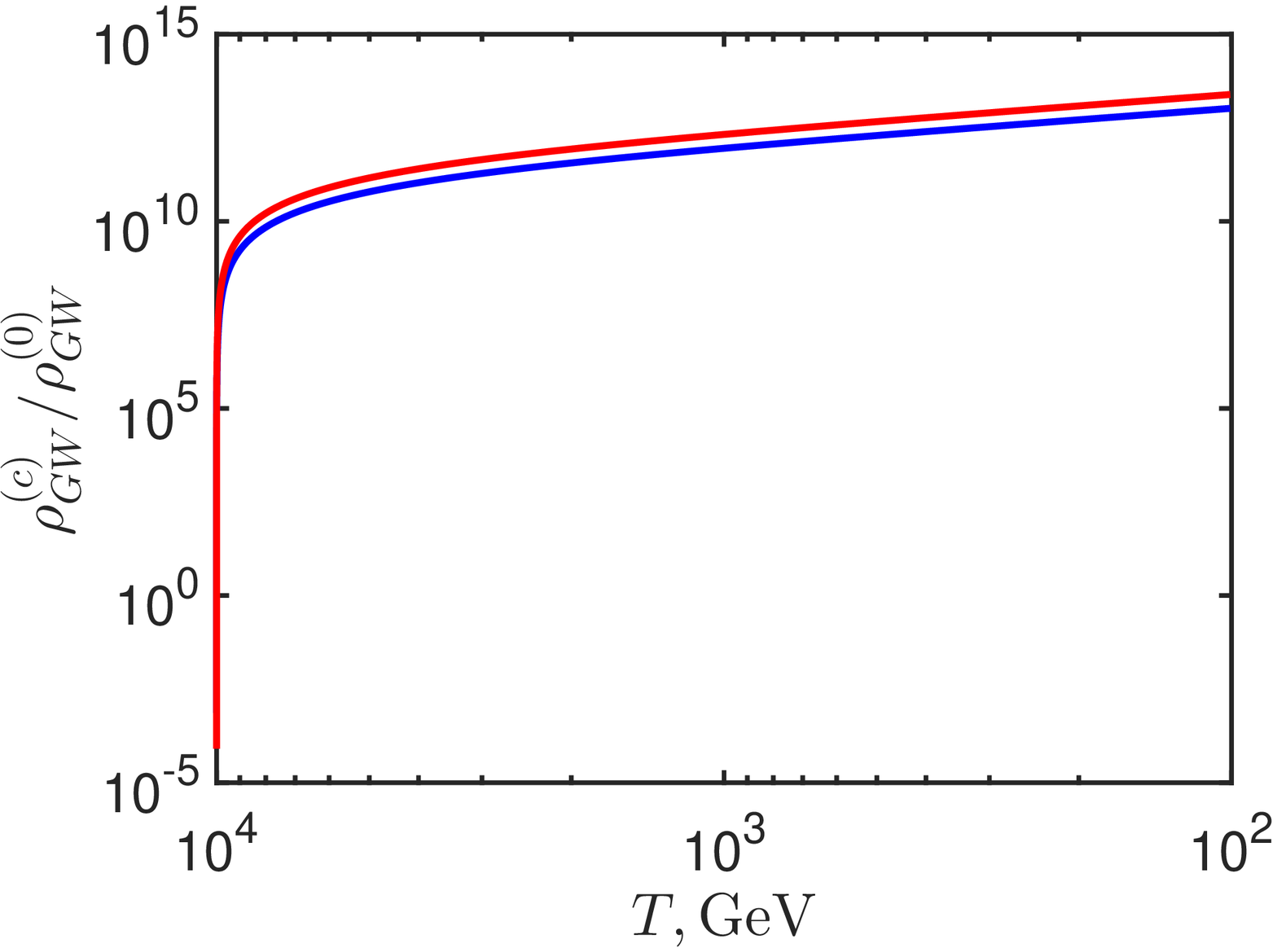}}
  \hskip-.5cm
  \subfigure[]
  {\label{fig:3f}
  \includegraphics[scale=.3]{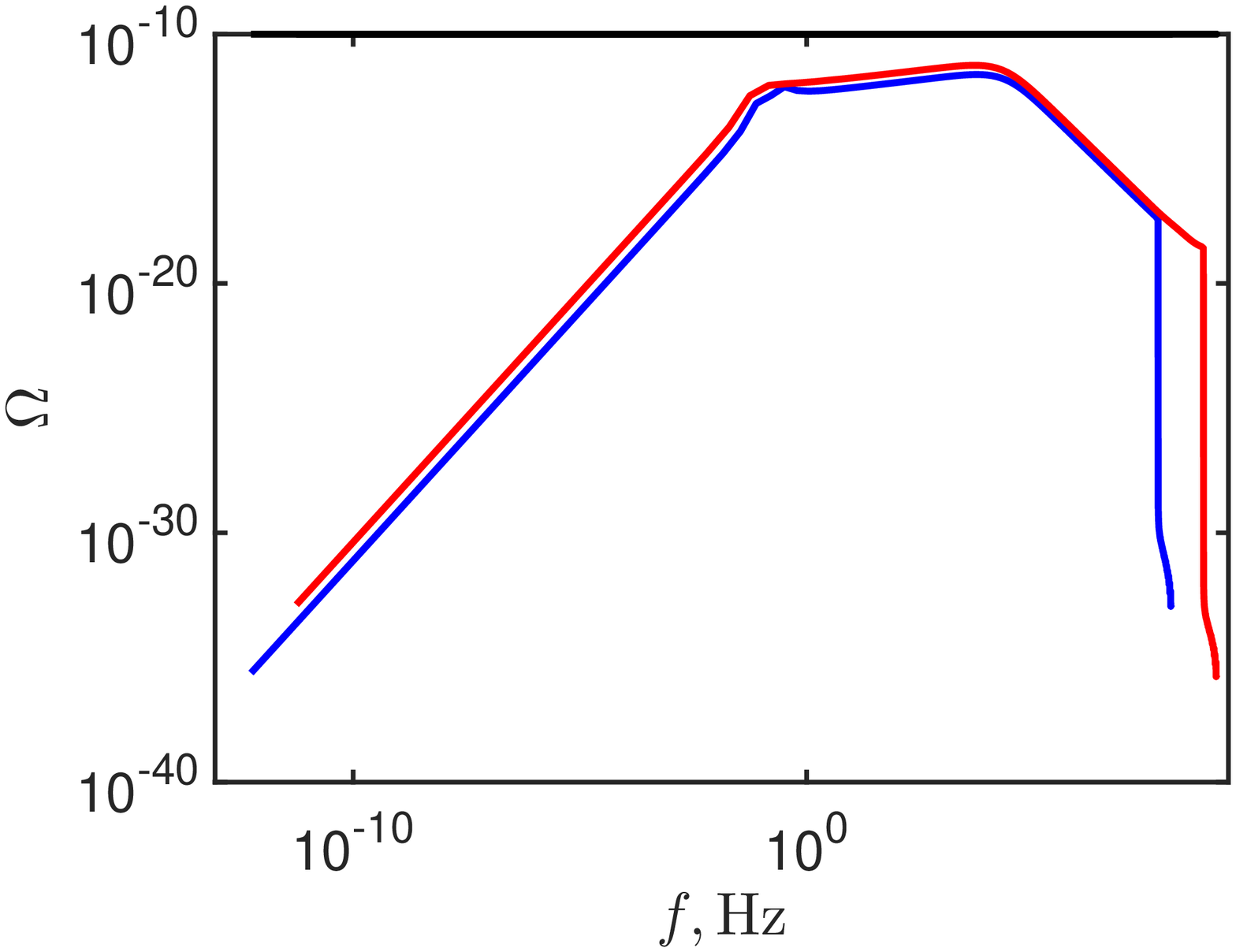}}
  \protect
  \vskip-.4cm
\caption{The conformal energy density of GWs $\rho_{\mathrm{GW}}^{(c)}$ in
Eq.~(\ref{eq:densGW}) normalized by $\rho_{\mathrm{GW}}^{(0)}$
in Eq.~(\ref{eq:rho0}) versus the plasma temperature $T$ is shown
in panels (a), (c), and (e) for different parameters of the system.
The spectral energy density of GWs in Eq.~(\ref{eq:specGWdmless}),
depending on the conformal frequency $f$, is depicted in panels (b),
(d), and (f) for different parameters of the system. In panels (b),
(d), and (f), by solid black line, we show the observational constraint
on $\Omega$, $\Omega_{\mathrm{obs}}\sim10^{-10}$, established in
Ref.~\citen{Abb21}. Blue line in panels (a) and (b) correspond to
$\tilde{B}_{\mathrm{Y}}^{(0)}=1.4\times10^{-2}$ and red one to $\tilde{B}_{\mathrm{Y}}^{(0)}=1.4\times10^{-1}$.
Additionally $\gamma_{\star}=10^{-2}$ and $\tilde{k}_{\mathrm{max}}=10^{-3}$
in panels (a) and (b). Blue line in panels (c) and (d) correspond
to $\gamma_{\star}=10^{-2}$ and red one to $\gamma_{\star}=10^{-3}$.
Additionally $\tilde{B}_{\mathrm{Y}}^{(0)}=1.4\times10^{-2}$ and
$\tilde{k}_{\mathrm{max}}=10^{-3}$ in panels (c) and (d). Blue line
in panels (e) and (f) correspond to $\tilde{k}_{\mathrm{max}}=10^{-3}$
and red one to $\tilde{k}_{\mathrm{max}}=10^{-2}$. Additionally
$\tilde{B}_{\mathrm{Y}}^{(0)}=1.4\times10^{-2}$ and $\gamma_{\star}=10^{-2}$
in panels (e) and (f).\label{fig:GW}}
\end{figure}

First, basing on the numerical solution of Eq.~(\ref{eq:HMFsys})
(see also Fig.~\ref{fig:spectra}), we show the evolution of $\rho_{\mathrm{GW}}^{(c)}(\eta)$
in the cooling universe in Figs.~\ref{fig:3a}, \ref{fig:3c}
and~\ref{fig:3e} normalized by
\begin{equation}\label{eq:rho0}
  \rho_{\mathrm{GW}}^{(0)}=
  \frac{\sigma_{c}^{2}\pi^{2}t_{\text{Univ}}^{2}T_{0}^{8}T_{\mathrm{RL}}^{2}G}{576\alpha'^{4}\tilde{M}_{\mathrm{Pl}}^{2}}=
  1.5\times10^{-23}\,\text{eV}\cdot\text{cm}^{-3},
\end{equation}
for different parameters of the system. We can see in Figs.~\ref{fig:3a},
\ref{fig:3c} and~\ref{fig:3e} that the energy density
of GWs grows despite the HMF strength diminishes; cf. Figs.~\ref{fig:2a},
\ref{fig:2c} and~\ref{fig:2e}. As explained in Ref.~\refcite{Dvo22},
it results from the cumulative integration over $\xi$ in Eq.~(\ref{eq:rhoGWisotr}).
Note that the behavior of $\rho_{\mathrm{GW}}^{(c)}(\eta)$ qualitatively
resembles that found in Ref.~\refcite{Bra21}.

We can see in Fig.~\ref{fig:3a} that the energy density of GWs
mainly depends on the seed strength $\tilde{B}_{\mathrm{Y}}^{(0)}$
since the integrand in Eq.~(\ref{eq:rhoGWisotr}) is quadratic in
the HMF spectra. The dependence on $\gamma_{\star}$ and $\tilde{k}_{\mathrm{max}}$
is not so pronounced. In general, the behavior of $\rho_{\mathrm{GW}}^{(c)}(\eta)$
for the seed spectrum in Eq.~\ref{eq:seedR0} resembles that for
the Kolmogorov seed spectrum, which was studied in Ref.~\refcite{Dvo22}.

We obtain the evolution of $\rho_{\mathrm{GW}}^{(c)}(\eta)$ produced by random HMFs accounting for the (H)MHD turbulence, i.e. we replace solving the Navier-Stokes equation for the plasma velocity by the consideration of the effective magnetic diffusion and the $\alpha$-dynamo parameters in Eq.~\eqref{eq:etaalphaY}. The production of relic GWs by fluctuating magnetic fields in the chiral plasma was recently studied in Ref.~\refcite{Bra21}, where the full set of the MHD equations was solved numerically. The behavior of the total energy density of GWs, found in Ref.~\refcite{Bra21}, is qualitatively the same as in Figs.~\ref{fig:3a}, \ref{fig:3c}, and~\ref{fig:3e}. It means that the approximation of the (H)MHD turbulence, applied to a chiral plasma, is valid. This fact was also mentioned in Ref.~\refcite{Dvo22}.

However, the current GW telescopes cannot measure the GW energy in
all range of frequencies. Instead, they are sensitive to the GW energy
density in a certain frequency interval. In particular, we discuss
the quantity
\begin{equation}\label{eq:Omegadef}
  \Omega(f)=\frac{f\rho_{\mathrm{GW}}(f)}{\rho_{\mathrm{crit}}},
\end{equation}
where $\rho_{\mathrm{crit}}=0.53\times10^{-5}\,\text{GeV}\cdot\text{cm}^{-3}$
is the critical energy density of the universe. The derivation of Eq.~\eqref{eq:Omegadef} is given in~\ref{sec:OMEGDER}. Since $k$ is the
conformal momentum, we can call $f=k/2\pi=5.6\times10^{10}\tilde{k}_{\mathrm{max}}\kappa\,\text{Hz}$
as the conformal frequency measured in Hz.

We can suppose that GWs generated by $T=T_{\mathrm{EW}}$ remains
unchanged at lower temperatures since HMFs are converted to Maxwell
ones. Thus, the spectral density at the present moment is $\rho_{\mathrm{GW}}(f)=\rho_{\mathrm{GW}}^{(c)}(f,\eta_{\mathrm{EW}})/a_{\mathrm{now}}^{3}=\rho_{\mathrm{GW}}^{(c)}(f,\eta_{\mathrm{EW}})$
since the present scale factor $a_{\mathrm{now}}=1$. Using Eqs.~(\ref{eq:rhoGWisotr})
and~(\ref{eq:newvar}), we rewrite Eq.~(\ref{eq:Omegadef}) in the
form,
\begin{align}\label{eq:specGWdmless}
  \Omega= & \Omega_{0}\frac{\tau_{\mathrm{EW}}}{\kappa^{2}}
  \int_{0}^{\tau_{\mathrm{EW}}}\frac{\mathrm{d}\tau'}{(1+7.1\times10^{-13}\tau'/\tilde{k}_{\mathrm{max}}^{2})^{2}}
  \nonumber
  \\
  & \times
  \iint_{S(\kappa)}\frac{\mathrm{d}\upsilon\mathrm{d}\varpi}{\upsilon{}^{3}\varpi^{3}}
  \nonumber
  \\
  & \times
  \big\{
    [4\kappa^{2}\upsilon^{2}+(\kappa^{2}+\upsilon^{2}-\varpi^{2})^{2}]
    [4\kappa^{2}\varpi^{2}+(\kappa^{2}-\upsilon^{2}+\varpi^{2})^{2}]R(\upsilon,\tau')R(\varpi,\tau')
    \nonumber
    \\
    & +
    16\kappa^{2}\upsilon^{2}\varpi^{2}(\kappa^{2}+\upsilon^{2}-\varpi^{2})(\kappa^{2}-\upsilon^{2}+\varpi^{2})
    H(\upsilon,\tau')H(\varpi,\tau')
  \big\},
\end{align}
where
\begin{equation}
  \Omega_{0}=\frac{\sigma_{c}^{2}\pi^{2}t_{\text{Univ}}^{2}T_{0}^{8}T_{\mathrm{RL}}^{2}G}
  {576\alpha'^{4}\tilde{M}_{\mathrm{Pl}}^{2}\rho_{\mathrm{crit}}}=1.6\times10^{-24}.
\end{equation}
Here, $S(\kappa)$ is the polygon integration domain in the $(\upsilon\varpi)$-plane.

We show $\Omega$ versus $f$ in Figs.~\ref{fig:3b}, \ref{fig:3d},
and~\ref{fig:3f} for different parameters of the system. We can
see that $\Omega$ reaches its maximal values in a plateau which spans
from $\sim\text{Hz}$ to $\sim\text{kHz}$. It matches the frequency
range probed in Ref.~\refcite{Abb21} in the search of a stochastic
GW background. The observational upper bound established in Ref.~\refcite{Abb21}
is $\Omega_{\mathrm{obs}}\sim10^{-10}$, which is also depicted in
Figs.~\ref{fig:3b}, \ref{fig:3d}, and~\ref{fig:3f}.
For the convenience, we summarize the parameters of the plateaus in
Table~\ref{tab:plat}.

\begin{table}[ph]
  \tbl{The frequency range and the values of $\Omega$ corresponding to the
  plateaus in Figs.~\ref{fig:3b}, \ref{fig:3d}, and~\ref{fig:3f}.}
  {
  \begin{tabular}{@{}llcc@{}} \toprule
    Figure & Type of line & $f_{\mathrm{plat}}$ ($f_{\mathrm{\mathrm{max}}}$) &
    $\Omega_\mathrm{plat}$ ($\Omega_{\mathrm{\mathrm{max}}}$) \\
    \colrule
    Fig.~\ref{fig:3b} ($\gamma_{\star}=10^{-2}$ &
    Blue line ($\tilde{B}_{\mathrm{Y}}^{(0)}=1.4\times10^{-2}$) &
    $0.3\,\text{Hz}<f_{\mathrm{plat}}<5.9\,\text{kHz}$ & 
    $7.9\times10^{-13}<\Omega_\mathrm{plat}<2.4\times10^{-12}$ \\ 
    and $\tilde{k}_{\mathrm{max}}=10^{-3}$) &
    Red line ($\tilde{B}_{\mathrm{Y}}^{(0)}=1.4\times10^{-1}$) &
    $0.3\,\text{Hz}<f_\mathrm{plat}<631\,\text{Hz}$ &
    $7.8\times10^{-9}<\Omega_\mathrm{plat}<1.5\times10^{-8}$ \\
    \cline{2-4} \cline{3-4} \cline{4-4}
    Fig.~\ref{fig:3d} ($\tilde{B}_{\mathrm{Y}}^{(0)}=1.4\times10^{-2}$  & 
    Blue line ($\gamma_{\star}=10^{-2}$) & 
    $0.3\,\text{Hz}<f_\mathrm{plat}<5.9\,\text{kHz}$ &   
    $7.9\times10^{-13}<\Omega_\mathrm{plat}<2.4\times10^{-12}$ \\
    and $\tilde{k}_{\mathrm{max}}=10^{-3}$) &
    Red line ($\gamma_{\star}=10^{-3}$) & $1.1\,\text{Hz}<f_\mathrm{plat}<5.9\,\text{kHz}$ &
    $9.5\times10^{-13}<\Omega_\mathrm{plat}<1.3\times10^{-12}$ \\
    \cline{2-4} \cline{3-4} \cline{4-4}
    Fig.~\ref{fig:3f} ($\tilde{B}_{\mathrm{Y}}^{(0)}=1.4\times10^{-2}$ & 
    Blue line ($\tilde{k}_{\mathrm{max}}=10^{-3}$) & 
    $0.3\,\text{Hz}<f_\mathrm{plat}<5.9\,\text{kHz}$ &   
    $7.9\times10^{-13}<\Omega_\mathrm{plat}<2.4\times10^{-12}$ \\ 
    and $\gamma_{\star}=10^{-2}$) &
    Red line ($\tilde{k}_{\mathrm{max}}=10^{-2}$) &   
    $f_{\mathrm{\mathrm{max}}}=4.6\,\text{kHz}$ &
    $\Omega_{\mathrm{\mathrm{max}}}=5.7\times10^{-12}$  \\ \botrule
  \end{tabular}\label{tab:plat}}
\end{table}

The plateau in Figs.~\ref{fig:3b}, \ref{fig:3d}, and~\ref{fig:3f}
is defined as the part of a curve between two first maxima in $\Omega$
if the difference between them is about one order of magnitude. If
the difference between maxima is greater, we deal with just a single
maximum, as in the last row in Table~\ref{tab:plat}.

We can see that only in the case corresponding to $\gamma_{\star}=10^{-2}$,
$\tilde{k}_{\mathrm{max}}=10^{-3}$, and $\tilde{B}_{\mathrm{Y}}^{(0)}=1.4\times10^{-1}$,
shown in Fig.~\ref{fig:3b} by the red line (see also the second
row in Table~\ref{tab:plat}), $\Omega$, predicted in our model,
exceeds $\Omega_{\mathrm{obs}}$. Thus, we can put the constraint
on $\tilde{B}_{\mathrm{Y}}^{(0)}$, $\tilde{B}_{\mathrm{Y}}^{(0)}\lesssim10^{-1}$,
or $B_{\mathrm{Y}}^{(0)}=T_{\mathrm{RL}}^{2}\tilde{B}_{\mathrm{Y}}^{(0)}\lesssim5\times10^{26}\,\text{G}$.
Such a HMF, while evolving down to the big bang nucleosynthesis (BBN)
temparature $T_{\mathrm{BBN}}=0.1\,\text{MeV}$, has the strength
$B(T=T_{\mathrm{BBN}})=T_{\mathrm{BBN}}^{2}\tilde{B}_{\mathrm{Y}}^{(0)}=5\times10^{10}\,\text{G}$.
This constraint is in agreement with the BBN upper bound $B_{\mathrm{BBN}}=10^{11}\,\text{G}$
obtained in Ref.~\refcite{CheSchTru94}. Other parameters of the system
do not violate the observational constraints on $\Omega$.

\section{Flavor oscillations of SN neutrinos in relic GWs\label{sec:NUFLOSC}}

In this section, we examine how relic GWs produced by HMFs, described
in Sec.~\ref{sec:PRODGWs}, influence flavor oscillations of SN neutrinos.
The interaction between stochastic GWs and neutrinos, as well as neutrino
flavor oscillations were studied in Refs.~\refcite{Dvo19,Dvo20,Dvo21}.

We consider the system of three mixed massive flavor neutrinos $\lambda=\nu_{e},\nu_{\mu},\nu_{\tau}$
interacting with stochastic GWs. The probability to detect a certain
neutrino flavor $\lambda=\nu_{e},\nu_{\mu},\nu_{\tau}$ in a neutrino
beam, which travels the distance $x$ between the emission and the
detection points, was found in Ref.~\refcite{Dvo21},
\begin{align}\label{eq:Plambdag}
  P_{\lambda}^{(g)}(x)= & \sum_{\sigma}P_{\sigma}(0)
  \bigg[
    \sum_{a}|U_{\lambda a}|^{2}|U_{\sigma a}|^{2}
    \notag
    \\
    & +
    2\text{Re}\sum_{a>b}U_{\lambda a}U_{\lambda b}^{*}U_{\sigma a}^{*}U_{\sigma b}\exp\left(-\mathrm{i}\varphi_{ab}x\right)g_{ab}
  \bigg],
\end{align}
where $P_{\sigma}(0)$ are the emission probabilities, which satisfy
$\sum_{\sigma}P_{\sigma}(0)=1$, $(U_{\lambda a})$ is the mixing
matrix between mass and flavor bases, $\varphi_{ab}=\frac{\Delta m_{ab}^{2}}{2E}$
are the phases of neutrino vacuum oscillations, $\Delta m_{ab}^{2}=m_{a}^{2}-m_{b}^{2}$,
with $a,b=1,2,3$, are the differences of the masses squared of the
neutrino mass eigenstates, $E$ is the mean neutrino energy, $g_{ab}(x)=\exp\left[-\varphi_{ab}^{2}\int_{0}^{x}g(t)\mathrm{d}t\right],$
$g(t)=\frac{3}{128}\int_{0}^{t}\mathrm{d}t_{1}\left(\left\langle h_{+}(t)h_{+}(t_{1})\right\rangle +\left\langle h_{\times}(t)h_{\times}(t_{1})\right\rangle \right)$,
and $h_{+,\times}$ are the random amplitudes of `plus' and `times'
polarizations of the GW background. The details of the derivation
of Eq.~(\ref{eq:Plambdag}) are also given in~\ref{sec:DENSMATR}.
The observed neutrino fluxes are $F_{\lambda}\propto P_{\lambda}^{(g)}$.

If we consider SN neutrinos, basing on Eq.~(\ref{eq:Plambdag}),
one gets that only solar oscillations channel contributes to the probabilities.
It is convenient to subtract the effect of neutrino vacuum oscillations
from the total probabilities by considering $\Delta P_{\lambda}=P_{\lambda}^{(g)}-P_{\lambda}^{(\mathrm{vac})}$.
Finally, we can rewrite $\Delta P_{\lambda}$ in the form~\cite{Dvo21},
\begin{align}\label{eq:DeltaP21}
  \Delta P_{\lambda}(x)= & 2
  \left[
    \text{Re}
    \left[
      U_{\lambda2}U_{\lambda1}^{*}U_{e2}^{*}U_{e1}
    \right]
    \cos
    \left(
      2\pi\frac{x}{L_{21}}
    \right)+
    \text{Im}
    \left[
      U_{\lambda2}U_{\lambda1}^{*}U_{e2}^{*}U_{e1}
    \right]
    \sin
    \left(
      2\pi\frac{x}{L_{21}}
    \right)
  \right]
  \nonumber
  \\
  & \times
  \left[
    1-\exp
    \left(
      -\Gamma_{\nu}
    \right)
  \right],
  \quad
  \Gamma_{\nu}=\frac{4\pi^{2}}{L_{21}^{2}}\int_{0}^{x}g(t)\mathrm{d}t.
\end{align}
where $L_{21}=\tfrac{4\pi E}{\Delta m_{21}^{2}}$ is the oscillations
length in vacuum for the solar channel. We should evaluate the parameter
$\Gamma_{\nu}$ for the GW background considered in Sec.~\ref{sec:PRODGWs}.
The contribution of GWs would be sizable if $\Gamma_{\nu}>1$.

We can rewrite the correlators of the amplitudes using the spectral
density $S(\omega)$ as
\begin{equation}\label{eq:specdens}
  \left\langle
    h_{+}(t)h_{+}(t_{1})
  \right\rangle +
  \left\langle
    h_{\times}(t)h_{\times}(t_{1})
  \right\rangle =
  \int_{0}^{\infty}\mathrm{d}\omega S(\omega)\cos[\omega(t-t_{1})].
\end{equation}
Thus,
\begin{equation}
  g(t)=\frac{3}{128}\int_{0}^{\infty}\frac{\mathrm{d}\omega}{\omega}\sin(\omega t)S(\omega),
\end{equation}
and
\begin{equation}\label{eq:Gammax}
  \Gamma_{\nu}(x)=\frac{6\pi^{3}G}{L_{21}^{2}}
  \int_{0}^{\infty}\frac{\mathrm{d}\omega}{\omega^{4}}\sin^{2}(\omega x/2)\rho_{\mathrm{GW}}(\omega),
\end{equation}
since $k=\omega$ for a GW wave. Hence, the spectrum of the energy density
is related to $S(\omega)$ by $\rho_{\mathrm{GW}}(k=\omega)=\omega^{2}S(\omega)/32\pi G$.
As a rule, the distance between the Earth, where we observe the neutrino
flux, and SN is great. Thus, we should consider Eq.~(\ref{eq:Gammax})
in the limit $x\to\infty$,
\begin{equation}\label{eq:GammaEarth}
  \Gamma_{\oplus}(x)=\frac{3\pi^{4}G}{4L_{21}^{2}}x\lim_{\omega\to0}\frac{\rho_{\mathrm{GW}}(\omega)}{\omega^{2}},
\end{equation}
where take into account the $\delta$-function definition, $\delta(\alpha)=\lim_{x\to\infty}\frac{\sin^{2}(\alpha x)}{\pi x\alpha^{2}}$.

The spectrum of the energy density of GWs is given in Eq.~(\ref{eq:rhoGWisotr}).
We suppose that there is no GW production after EWPT, i.e. the conformal
spectrum is constant after EWPT. Thus we should put $\eta=\eta_{\mathrm{EW}}$
in Eq.~(\ref{eq:rhoGWisotr}). Considering the limit $k\to0$ and
taking that $\rho_{\mathrm{GW}}(\omega)=\rho_{\mathrm{GW}}^{(c)}(k,\eta_{\mathrm{EW}})$
since we study the neutrino propagation and oscillations at the present
time, one gets that
\begin{align}
  \rho_{\mathrm{GW}}^{(c)}(k,\eta_{\mathrm{EW}})\to k^{2} & \frac{8t_{\text{Univ}}^{2}G}{\pi^{2}}\eta_{\mathrm{EW}}
  \int_{0}^{\eta_{\mathrm{EW}}}\frac{\mathrm{d}\xi}{(\eta_{0}+\xi)^{2}}\int_{0}^{k_{\text{max}}}\frac{\mathrm{d}p}{p^{2}}
  \left[
    \rho_{\mathrm{Y}}^{(c)}(p,\xi)
  \right]^{2}.
\end{align}
Finally, using Eq.~(\ref{eq:newvar}), we rewrite Eq.~(\ref{eq:GammaEarth})
as
\begin{equation}\label{eq:Gamma0}
  \Gamma_{\oplus}(x) =
  \left(
    \frac{x}{L}
  \right)
  \Gamma_{0},
  \quad
  \Gamma_{0}=NI_{\nu},
\end{equation}
where
\begin{align}\label{eq:NInu}
  N & =\frac{\sigma_{c}\pi^{6}t_{\text{Univ}}^{2}G^{2}T_{0}^{5}T_{\mathrm{RL}}L}
  {6\alpha'^{4}\tilde{M}_{\mathrm{Pl}}L_{21}^{2}\tilde{k}_{\mathrm{max}}^{3}}=
  \frac{1.6\times10^{-76}}{\tilde{k}_{\mathrm{max}}^{3}},
  \nonumber
  \\
  I_{\nu} & =\int_{0}^{\tau_{\mathrm{EW}}}
  \frac{\tau_{\mathrm{EW}}\mathrm{d}\tau}{(1+7.1\times10^{-13}\tau/\tilde{k}_{\mathrm{max}}^{2})^{2}}
  \int_{0}^{1}\frac{\mathrm{d}\kappa}{\kappa^{2}}R^{2}(\kappa,\tau).
\end{align}
Here we normalize the neutrino propagation distance on the typical
galaxy size $L=10\,\text{kpc}.$ The integral in Eq.~(\ref{eq:NInu})
is computed using the results of Sec.~\ref{subsec:NUMHMF} (see Fig.~\ref{fig:spectra}).

\begin{figure}
  \centering
  \includegraphics[scale=0.33]{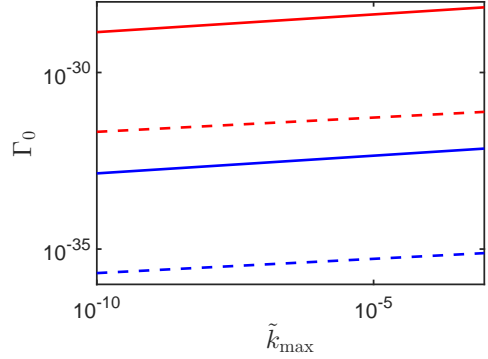}
  \protect
\caption{The parameter $\Gamma_{0}$ in Eq.~(\ref{eq:Gamma0}) versus $\tilde{k}_{\mathrm{max}}$
for different values of $\tilde{B}_{\mathrm{Y}}^{(0)}$ and $\gamma_{\star}$.
Red lines correspond to $\tilde{B}_{\mathrm{Y}}^{(0)}=1.4\times10^{-2}$
and blue ones to $\tilde{B}_{\mathrm{Y}}^{(0)}=1.4\times10^{-3}$.
Solid lines are plotted for $\gamma_{\star}=10^{2}$ and dashed ones
for $\gamma_{\star}=10^{3}$.\label{fig:Gamma0}}
\end{figure}

We show the parameter $\Gamma_{0}$ versus $\tilde{k}_{\mathrm{max}}$
for different $\tilde{B}_{\mathrm{Y}}^{(0)}$ and $\gamma_{\star}$
in Fig.~\ref{fig:Gamma0}. One can see that $\Gamma_{0}\ll1$ for
the reasonable values of all parameters. Even if we consider an extragalactic
SN and take $x\sim1\,\text{Gpc}$, $\Gamma_{\oplus}$ still remains quite
small. Thus, using Eq.~(\ref{eq:DeltaP21}), one gets that $\Delta P_{\lambda}\ll1$.

It means that the GW background generated by HMFs does not influence
flavor oscillations of SN neutrinos unlike stochastic GWs produced
by merging supermassive black holes (SMBHs), which were studied in
Ref.~\refcite{Dvo21}. Such a difference between the results is because
of the fact that $\Omega$ in case of stochastic GWs generated by
coalescing SMBHs is nonzero for $f\geq f_{\mathrm{min}}\neq0$. In
our case, $\rho_{\mathrm{GW}}(f)\propto f^{2}$ for small frequencies,
i.e. $f_{\mathrm{min}}=0$. 

\section{BAU caused by the lepton asymmetries\label{sec:BAU}}

The self-consistent evolution of HMFs and the asymmetries of leptons
and Higgs bosons in Eq.~(\ref{eq:HMFsys}) give rise to not only
relic GWs. If we focus on the $\xi_{e\mathrm{R,L}}$ evolution, we
can predict the generated BAU at EWPT. We studied this problem in
Ref.~\refcite{DvoSem21} basing on the Kolmogorov spectrum of seed HMFs.
Now we reexamine the issue of the BAU generation in HMFs using a more
realistic seed spectrum in Eq.~(\ref{eq:seedspecE}).

Given the asymmetries of right and left leptons $\xi_{e\mathrm{R,L}}$, BAU has the form~\cite{DvoSem21},
\begin{align}\label{eq:BAUgen}
  \text{BAU}(\tilde{\eta})= & \frac{n_{\mathrm{B}}-n_{\bar{\mathrm{B}}}}{s}=
  5.3\times10^{-3}\int_{0}^{\tilde{\eta}}\mathrm{d}\tilde{\eta}'
  \notag
  \\
  & \times
  \left\{
    \frac{{\rm d}\xi_{e\mathrm{R}}(\tilde{\eta}')}{{\rm d}\tilde{\eta}'}+
    \Gamma(\tilde{\eta}')[\xi_{e\mathrm{R}}(\tilde{\eta}')-\xi_{e\mathrm{L}}(\tilde{\eta}')]
  \right\} -
  \frac{6\times10^{7}}{\tilde{\eta}_{\mathrm{EW}}}
  \int_{0}^{\tilde{\eta}}\xi_{e\mathrm{L}}(\tilde{\eta}')\mathrm{d}\tilde{\eta}',
\end{align}
where $n_{\mathrm{B},\bar{\mathrm{B}}}$ are number densities of baryons
and antibaryons, and $s$ is the entropy density. Equation~\eqref{eq:BAUgen}
implies that $\text{BAU}=0$ at $T=T_{\mathrm{RL}}$.

\begin{figure}
  \centering
  \subfigure[]
  {\label{fig:5a}
  \includegraphics[scale=.3]{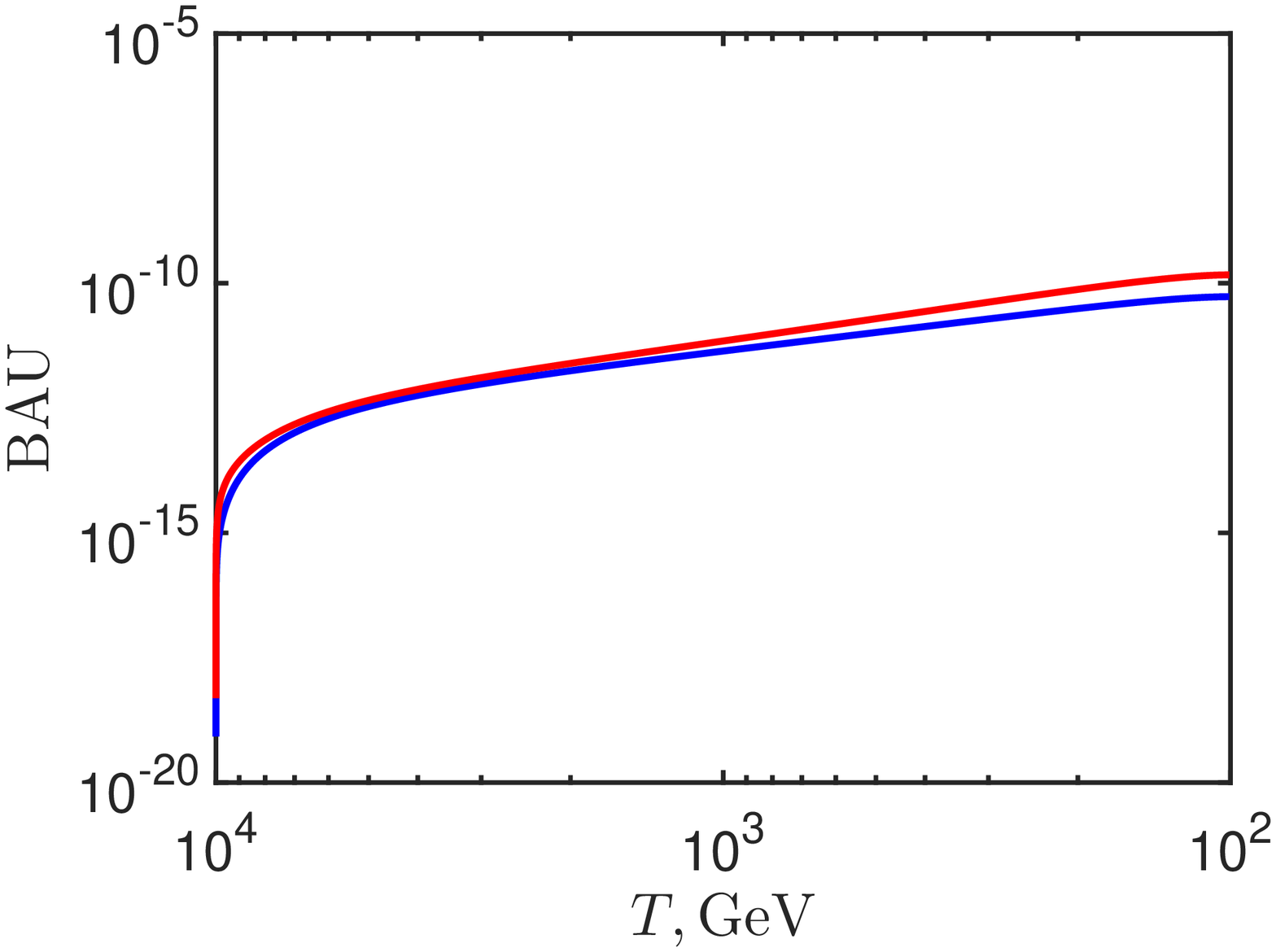}}
  \hskip-.5cm
  \subfigure[]
  {\label{fig:5b}
  \includegraphics[scale=.3]{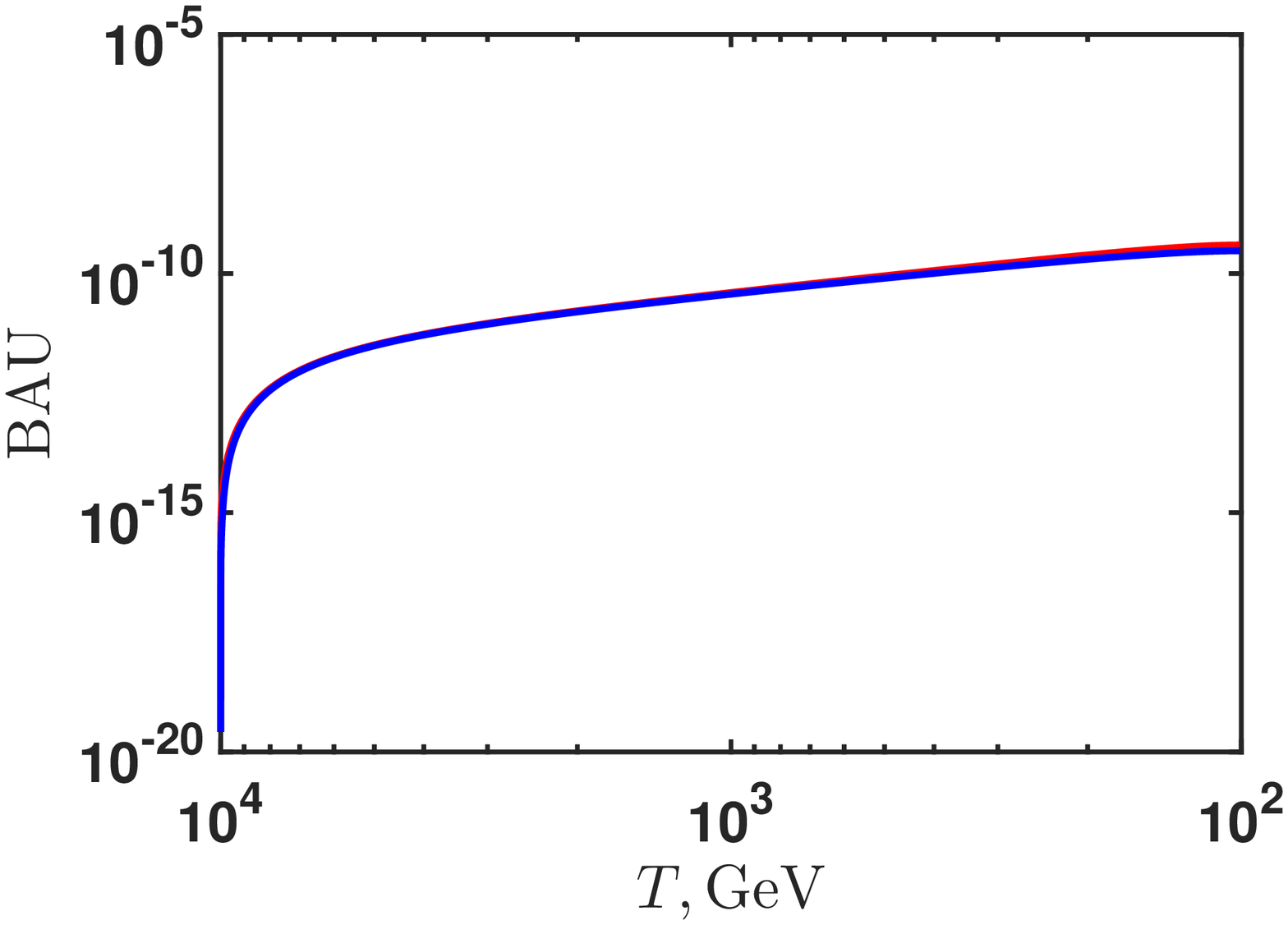}}
  \protect
\caption{The evolution of BAU based on the numerical solution of Eq.~(\ref{eq:HMFsys})
corresponding to the initial condition in Sec.~\ref{subsec:SEEDSPEC}.
We take that $\tilde{B}_{\mathrm{Y}}^{(0)}=1.4\times10^{-6}$ and
$\tilde{k}_{\mathrm{max}}=10^{-3}$. Blue lines correspond to $\gamma_{\star}=10^{-2}$
and red ones to $\gamma_{\star}=10^{-3}$.
(a) $\xi_{e\mathrm{R}}^{(0)} = 10^{-10}$; (b) $\xi_{e\mathrm{R}}^{(0)} = 10^{-9}$.\label{fig:BAU}}
\end{figure}

We show the evolution of BAU in the universe cooling from $T_{\mathrm{RL}}$
down to $T_{\mathrm{EW}}$ for different $\gamma_{\star}$ and $\xi_{e\mathrm{R}}^{(0)}$ in Fig.~\ref{fig:BAU}.
For fixed $\tilde{B}_{\mathrm{Y}}^{(0)}$ and $\gamma_{\star}$, BAU
does not significantly depend on $\tilde{k}_{\mathrm{max}}$. That
is why we use the constant value $\tilde{k}_{\mathrm{max}}=10^{-3}$
in our simulations. If $\tilde{B}_{\mathrm{Y}}^{(0)}>1.4\times10^{-6}$,
BAU exceeds the observed value $\text{BAU}_{\mathrm{obs}}\sim10^{-10}$.
Thus, we do not consider quite strong seed HMFs as we made in Secs.~\ref{subsec:NUMHMF}
and~\ref{subsec:RESGWs}.

In Figs.~\ref{fig:5a} and~\ref{fig:5b}, we also examine the dependence of BAU on the initial right electrons asymmetry. When $\xi_{e\mathrm{R}}^{(0)} = 10^{-10}$, we get that $\text{BAU}_\mathrm{obs} \sim 10^{-10}$ at EWPT. The enhancement of $\xi_{e\mathrm{R}}^{(0)}$ results in BAU which exceeds the observed value; cf Fig.~\ref{fig:5b}. The excessive BAU in Fig.~\ref{fig:5b} cannot be reduced by varying other parameters of the system. Thus, the generated BAU is almost completely defined by the initial right electrons asymmetry.

One can see in Fig.~\ref{fig:5a} that we reach the observed BAU
at $T=T_{\mathrm{EW}}=10^{2}\,\text{GeV}$ if $\tilde{B}_{\mathrm{Y}}^{(0)}=1.4\times10^{-6}$.
This result corresponds to the seed spectrum in Eq.~(\ref{eq:seedspecE}).
We obtained in Ref.~\refcite{DvoSem21} that $\text{BAU}_{\mathrm{obs}}\sim10^{-10}$
can be achieved at significantly stronger seed HMFs with $\tilde{B}_{\mathrm{Y}}^{(0)}\sim1.4\times(10^{-2}\div10^{-1})$.
However, those HMFs in Ref.~\refcite{DvoSem21} corresponded to the
seed Kolmogorov spectrum. It means that the consideration of BAU as
a consequence of the evolution of HMFs and the asymmetries in Eq.~(\ref{eq:HMFsys})
imposes a stronger constraint on $\tilde{B}_{\mathrm{Y}}^{(0)}$ in
comparison with the relic GWs background studied in Sec.~\ref{subsec:RESGWs}.

\section{Conclusion\label{sec:CONCL}}

In the present work, we have studied the evolution of random HMFs
in the early universe cooling down from $T_{\mathrm{RL}}=10\,\mathrm{TeV}$
to $T_{\mathrm{EW}}=100\,\mathrm{GeV}$. The evolution of HMFs is
driven by the analog of the CME and accounts for the (H)MHD turbulence.
The analog of the CME involves the asymmetries, proportional to the chiral
imbalances, of right and left leptons.
The initial temperature $T_{\mathrm{RL}}=10\,\mathrm{TeV}$ is chosen
so that the contributions of the asymmetries in Eq.~(\ref{eq:HMFsys})
are self-consistent, i.e. left leptons start to be produced. The final
temperature $T_{\mathrm{EW}}=100\,\mathrm{GeV}$ corresponds to EWPT
when particles acquire masses and the chiral approximation is no longer
valid.

The (H)MHD turbulence implies the dominant role of the Lorentz force
in the Navier-Stokes equation. In~\ref{sec:MHDTURB}, we
have rederived the turbulent terms in the kinetic equations for the spectra
of the magnetic energy and the magnetic helicity (see Eqs.~(\ref{eq:etaalphaY})
and~(\ref{eq:etaalpha})). We have corrected the form of the $\alpha$-dynamo
parameter used in Refs.~\refcite{DvoSem21,Dvo22,DvoSem17}. Now, our
results are consistent with Ref.~\refcite{Cam07}. However, as found
in Sec.~\ref{subsec:NUMHMF} and claimed in Refs.~\refcite{DvoSem21,Dvo22},
the evolution of HMFs is dominated by the diffusion rather than the
$\alpha$-dynamo term.

The main advance of the present work in comparison with Refs.~\refcite{DvoSem21,Dvo22}
is the consideration of a more realistic seed spectrum of HMFs which
is Batchelor at small momenta and Kolmogorov at great ones. The necessity
of the vanishing spectrum at great length scales, comparable with
the horizon size, was mentioned in Ref.~\refcite{Bra17}. Now, we do
not have to consider the minimal momentum. However, we should take
into account the border momentum $\tilde{k}_{\star}$ (see Eq.~(\ref{eq:seedspecE})),
or $\gamma_{\star}$, which is a new free parameter in the system.

In Sec.~\ref{subsec:SEEDSPEC}, we have formulated the initial condition
for Eq.~(\ref{eq:HMFsys}) and numerically solved it in Sec.~\ref{subsec:NUMHMF}.
The system in Eq.~(\ref{eq:HMFsys}) has been represented in the
form convenient for numerical simulations in~\ref{sec:NEWVAR}.
We have obtained the behavior of the spectra of the magnetic energy
and the magnetic helicity, as well as the evolution the HMFs strength
and the $\alpha$-dynamo parameter; cf. Figs.~\ref{fig:spectra}
and~\ref{fig:B}. Qualitatively, the evolution of these parameters
resembles that found in Refs.~\refcite{DvoSem21,Dvo22}.

We have studied the various phenomena affected by HMFs. First, in
Sec.~\ref{fig:GW}, we have considered the production of relic GWs
by random HMFs. For this purpose, we have used the formalism developed
in Ref.~\refcite{Dvo22}. We have tracked the evolution of the energy
density of GWs from $T_{\mathrm{RL}}$ down to EWPT in Sec.~\ref{subsec:RESGWs}.
We have also discussed the observability of the predicted GW background
by the current GW detectors. The maximum of the spectral density of the
predicted GW signal is in the range from $\sim\text{Hz}$ to $\sim\text{kHz}$.
It coincides with the sensitivity of the LIGO-Virgo-KAGRA collaborations~\cite{Abb21}.
Thus, in Sec.~\ref{subsec:RESGWs}, we could establish the constraint
on the strength of HMF which is $\tilde{B}_{\mathrm{Y}}^{(0)}\lesssim10^{-1}$
or $B_{\mathrm{Y}}^{(0)}<5\times10^{26}\,\text{G}$. The obtained
upper limit is consistent with the BBN constraint on the magnetic field
strength derived in Ref.~\refcite{CheSchTru94}.

In Sec.~\ref{sec:NUFLOSC}, we have discussed flavor oscillations
of SN neutrinos in relic GWs predicted in our model. We have used
the formalism for the description of neutrino flavor oscillations
in GWs developed in Refs.~\refcite{Dvo19,Dvo20,Dvo21}. Some of the
issues of this formalism were clarified in~\ref{sec:DENSMATR}.
We have analyzed whether the neutrino interaction with GWs can modify
the observed fluxes of SN neutrinos. We have obtained that the contribution
of GWs to the fluxes is rather small for realistic SN neutrinos.

Finally, in Sec.~\ref{sec:BAU}, we have calculated BAU which is
generated in the wake of the evolution of the asymmetries of right
and left leptons. For this purpose, we
have applied the technique developed in Ref.~\refcite{DvoSem21}. We
have obtained that, in order not to exceed the observed $\text{BAU}_{\mathrm{obs}}\sim10^{-10}$,
we should constrain the strength of the seed HMF by $\tilde{B}_{\mathrm{Y}}^{(0)}\lesssim10^{-6}$,
or $B_{\mathrm{Y}}^{(0)}<5\times10^{20}\,\text{G}$. This upper limit
is stronger than that derived in Sec.~\ref{subsec:RESGWs} basing
on the observability of relic GWs. Moreover, it is also stronger than
the result of Ref.~\refcite{DvoSem21}, where analogous problem was
studied and we used the Kolmogorov seed spectrum. The upper bound on the seed HMF derived from the consideration of BAU is consistent with the result of Ref.~\refcite{Kam20}.

Generally, the $\alpha$-dynamo mechanism used in our work to drive the evolution of HMFs requires a nonzero seed field $B_{\mathrm{Y}}^{(0)}$. We do not explain origin of $B_{\mathrm{Y}}^{(0)}$. Only constraints on the seed field are established. Nevertheless, there are models for the production of a seed field in the inflationary epoch (see, e.g., Ref.~\refcite{AnbSab15}). This inflation based HMF evolves in the cooling universe down to $T_\mathrm{RL} = 10\,\text{TeV}$ leading to $B_{\mathrm{Y}}^{(0)}$ used in our work. Quantum fluctuations during the inflation can be of the tensor type and lead to the production of GWs~\cite{Guz16}. In particular, we mention the generation of relic GWs in (pre-)inflationary times within the modifications of the General Relativity which was studied in Refs.~\refcite{Oik22a,Oik22b,Oik22c}. Various models for the production of primordial GWs in modified gravity theories, including the analysis of the GW spectra generated, are reviewed in Ref.~\refcite{OdiOikMyr22}.
There are attempts to probe such GWs~\cite{Ade21,Auc22}. In our work, we do not take into account primordial GWs produced during the inflation. The only possible impact of physics processes in the inflationary epoch on our results is the generation of a seed HMF.

\section*{Acknowledgments}

I am thankful to V.~B.~Semikoz and M.~E.~Shaposhnikov for the communications.

\appendix

\section{Contribution of the MHD turbulence to the coefficients in the kinetic
equations\label{sec:MHDTURB}}

In this appendix, we reexamine the contribution of random magnetic
fields to the kinetic equations for the magnetic energy and the helicity
within the approximation of the MHD turbulence. This problem was studied
in Refs.~\refcite{Cam07,DvoSem17} leading to the contradictory results.
We omit the subscript `Y' for brevity dealing with Maxwell magnetic fields.
The generalization of the results to HMFs is straightforward.

We shall keep the notations similar to Refs.~\refcite{Cam07,DvoSem17}
as close as possible. First, we fix the Fourier transform as $\mathbf{B}(\mathbf{k})=\int\mathrm{d}^{3}xe^{\mathrm{i}\mathbf{kx}}\mathbf{B}(\mathbf{x}).$
Since we study random fields, we take that the equal times correlator
of the magnetic fields strengths reads~\cite{Cam07,DvoSem17}
\begin{equation}\label{eq:correl}
  \left\langle
    B_{j}(\mathbf{k},t)B_{i}(\mathbf{p},t)
  \right\rangle =
  \frac{(2\pi)^{3}}{2}\delta(\mathbf{k}+\mathbf{p})
  \left[
    \left(
      \delta_{ji}-\hat{k}_{j}\hat{k}_{i}
    \right)
    S(k,t)+\mathrm{i}\varepsilon_{jin}\hat{k}_{n}A(k,t)
  \right],
\end{equation}
where $\hat{\mathbf{k}}=\mathbf{k}/k$, $k=|\mathbf{k}|$, and
\begin{equation}\label{eq:SAdef}
  S(k,t)=\frac{4\pi^{2}\rho(k,t)}{k^{2}},\quad A(k,t)=\frac{2\pi^{2}h(k,t)}{k},
\end{equation}
are related to the spectra of the densities of the magnetic energy
$\rho(k,t)$ and the helicity $h(k,t)$. The densities of energy and
helicity have the form, $B^{2}/2=\smallint\mathrm{d}k\rho(k,t)$ and
$h\equiv\smallint\mathrm{d}^{3}x(\mathbf{AB})/V=\smallint\mathrm{d}kh(k,t)$.
Here, we are in frames of the mean field approximation.

We use the MHD approximation~\cite{Sig02}, in which the plasma velocity is
$\mathbf{v}=\tfrac{\tau_{d}}{P+\rho}(\mathbf{J}\times\mathbf{B})$,
where $\tau_{d}$ is the phenomenological drag time, $\mathbf{J}$
is the electric current, $P$ is the plasma pressure, $\rho$ is the
energy density of matter. Substituting such velocity to the induction
equation, $\dot{\mathbf{B}}=\nabla\times(\mathbf{v}\times\mathbf{B})+\eta_{m}\Delta\mathbf{B}$,
where $\eta_{m}$ is the magnetic diffusion coefficient, and making
the Fourier transform, we derive the evolution equation in the form~\cite{Cam07,DvoSem17},
\begin{equation}\label{eq:evolXi}
  \dot{B}_{j}(\mathbf{k})=-\eta_{m}k^{2}B_{j}(\mathbf{k})+\Xi_{j}(\mathbf{k}),
\end{equation}
where
\begin{align}\label{eq:Xi}
  \Xi_{j}(\mathbf{k})= & \frac{\tau_{d}}{P+\rho}\int\frac{\mathrm{d}^{3}l}{(2\pi)^{3}}\frac{\mathrm{d}^{3}q}{(2\pi)^{3}}
  \varepsilon_{jtk}k_{t}q_{r}B_{s}(\mathbf{q})
  \notag
  \\
  & \times
  \left[
    \varepsilon_{krs}B_{n}(\mathbf{l}-\mathbf{q})B_{n}(\mathbf{k}-\mathbf{l})-
    \varepsilon_{rms}B_{k}(\mathbf{k}-\mathbf{l})B_{m}(\mathbf{l}-\mathbf{q})
  \right].
\end{align}
We differentiate Eq.~(\ref{eq:correl}) by time and use Eq.~(\ref{eq:evolXi}),
\begin{multline}\label{eq:correldiff}
  \left\langle
    \dot{B}_{j}(\mathbf{k},t)B_{i}(\mathbf{p},t)+B_{j}(\mathbf{k},t)\dot{B}_{i}(\mathbf{p},t)
  \right\rangle =
  \frac{(2\pi)^{3}}{2}\delta(\mathbf{k}+\mathbf{p})
  \\
  \times
  \left[
    \left(
      \delta_{ji}-\hat{k}_{j}\hat{k}_{i}
    \right)
    \dot{S}(k,t)+\mathrm{i}\varepsilon_{jin}\hat{k}_{n}\dot{A}(k,t)
  \right]
  \\ =
  -\eta_{m}k^{2}(2\pi)^{3}\delta(\mathbf{k}+\mathbf{p})
  \left[
    \left(
      \delta_{ji}-\hat{k}_{j}\hat{k}_{i}
    \right)
    S(k,t)+\mathrm{i}\varepsilon_{jin}\hat{k}_{n}A(k,t)
  \right]
  \\
  +
  \left\langle
    B_{i}(\mathbf{p},t)\Xi_{j}(\mathbf{k},t)+B_{j}(\mathbf{k},t)\Xi_{i}(\mathbf{p},t)
  \right\rangle.
\end{multline}
Multiplying Eq.~(\ref{eq:correldiff}) by $\delta_{ij}$ and $\mathrm{i}\varepsilon_{ijn}\hat{k}_{n}$,
we get the following equations for $S$ and $A$:
\begin{align}\label{eq:sysraw}
  \delta(\mathbf{k}+\mathbf{p})
  \left[
    \dot{S}(k,t)+2\eta_{m}k^{2}S(k,t)
  \right] & =
  \frac{1}{(2\pi)^{3}}
  \left\langle
    B_{i}(\mathbf{p},t)\Xi_{i}(\mathbf{k},t)+B_{i}(\mathbf{k},t)\Xi_{i}(\mathbf{p},t)
  \right\rangle,
  \nonumber
  \\
  \delta(\mathbf{k}+\mathbf{p})
  \left[
    \dot{A}(k,t)+2\eta_{m}k^{2}A(k,t)
  \right] & =
  \frac{\mathrm{i}\varepsilon_{ijn}\hat{k}_{n}}{(2\pi)^{3}}
  \left\langle
    B_{i}(\mathbf{p},t)\Xi_{j}(\mathbf{k},t)+B_{j}(\mathbf{k},t)\Xi_{i}(\mathbf{p},t)
  \right\rangle.
\end{align}
The quantity $\left\langle B_{i}(\mathbf{p},t)\Xi_{j}(\mathbf{k},t)+B_{j}(\mathbf{k},t)\Xi_{i}(\mathbf{p},t)\right\rangle $
contains four-point correlators which can be expressed through two-point
ones, as prescribed in Refs.~\refcite{Cam07,DvoSem17}.

Using Eq.~(\ref{eq:Xi}), we obtain that
\begin{align}\label{eq:corrXi}
  &
  \left\langle
    B_{i}(\mathbf{p})\Xi_{j}(\mathbf{k})+ B_{j}(\mathbf{k})\Xi_{i}(\mathbf{p})
  \right\rangle =
  \delta(\mathbf{k}+\mathbf{p})\frac{\tau_{d}}{4(P+\rho)}\int\mathrm{d}^{3}q
  \Big[
    2k_{t}k_{r}\varepsilon_{krs}S(q)
    \notag
    \displaybreak[2]
    \\
    & \times
    \Big\{
      \varepsilon_{jtk}
      \left[
        \left(
          \delta_{is}-\hat{k}_{i}\hat{k}_{s}
        \right)
        S(k)-\mathrm{i}\varepsilon_{isl}\hat{k}_{l}A(k)
      \right] +
      \varepsilon_{itk}
      \left[
        \left(
          \delta_{js}-\hat{k}_{j}\hat{k}_{s}
        \right)
        S(k)+\mathrm{i}\varepsilon_{jsl}\hat{k}_{l}A(k)
      \right]
    \Big\}
    \nonumber
    \displaybreak[2]
    \\
    & +
    2k_{t}q_{r}\varepsilon_{krs}
    \left[
      \left(
        \delta_{sn}-\hat{q}_{s}\hat{q}_{n}
      \right)
      S(q)+\mathrm{i}\varepsilon_{snl}\hat{q}_{l}A(q)
    \right]
    \notag
    \displaybreak[2]
    \\
    & \times    
    \Big\{
      \varepsilon_{jtk}
      \left[
        \left(
          \delta_{in}-\hat{k}_{i}\hat{k}_{n}
        \right)
        S(k)-\mathrm{i}\varepsilon_{inl}\hat{k}_{l}A(k)
      \right] -      
      \varepsilon_{itk}
      \left[
        \left(
          \delta_{jn}-\hat{k}_{j}\hat{k}_{n}
        \right)
        S(k)+\mathrm{i}\varepsilon_{jnl}\hat{k}_{l}A(k)
      \right]
    \Big\}
    \nonumber
    \displaybreak[2]
    \\
    & -
    k_{t}k_{r}\varepsilon_{rms}
    \left[
      \left(
        \delta_{km}-\hat{q}_{k}\hat{q}_{m}
      \right)
      S(q)-\mathrm{i}\varepsilon_{kml}\hat{q}_{l}A(q)
    \right]
    \notag
    \displaybreak[2]
    \\
    & \times    
    \Big\{
      \varepsilon_{jtk}
      \left[
        \left(
          \delta_{is}-\hat{k}_{i}\hat{k}_{s}
        \right)
        S(k)-\mathrm{i}\varepsilon_{isl}\hat{k}_{l}A(k)
      \right] +      
      \varepsilon_{itk}
      \left[
        \left(
          \delta_{js}-\hat{k}_{j}\hat{k}_{s}
        \right)
        S(k)+\mathrm{i}\varepsilon_{jsl}\hat{k}_{l}A(k)
      \right]
    \Big\}
    \nonumber
    \displaybreak[2]
    \\
    & -
    k_{t}q_{r}\varepsilon_{rms}
    \left[
      \left(
        \delta_{sm}-\hat{q}_{s}\hat{q}_{m}
      \right)
      S(q)+\mathrm{i}\varepsilon_{sml}\hat{q}_{l}A(q)
    \right]
    \notag
    \\
    & \times
    \Big\{
      \varepsilon_{jtk}
      \left[
        \left(
          \delta_{ik}-\hat{k}_{i}\hat{k}_{k}
        \right)
        S(k)-\mathrm{i}\varepsilon_{ikl}\hat{k}_{l}A(k)
      \right] -      
      \varepsilon_{itk}
      \left[
        \left(
          \delta_{jk}-\hat{k}_{j}\hat{k}_{k}
        \right)
        S(k)+\mathrm{i}\varepsilon_{jkl}\hat{k}_{l}A(k)
      \right]
    \Big\}
    \nonumber
    \displaybreak[2]
    \\
    & -
    k_{t}q_{r}\varepsilon_{rms}
    \left[
      \left(
        \delta_{sk}-\hat{q}_{s}\hat{q}_{k}
      \right)
      S(q)+\mathrm{i}\varepsilon_{skl}\hat{q}_{l}A(q)
    \right]
    \notag
    \\
    & \times    
    \Big\{
      \varepsilon_{jtk}
      \left[
        \left(
          \delta_{im}-\hat{k}_{i}\hat{k}_{m}
        \right)
        S(k)-\mathrm{i}\varepsilon_{iml}\hat{k}_{l}A(k)
      \right] -
      \varepsilon_{itk}
      \left[
        \left(
          \delta_{jm}-\hat{k}_{j}\hat{k}_{m}
        \right)
        S(k)+\mathrm{i}\varepsilon_{jml}\hat{k}_{l}A(k)
      \right]
    \Big\}
  \Big].
\end{align}
Equation~(\ref{eq:corrXi}) is written down in such a form intentionally.
It allows one to separate symmetric and antisymmetric combinations
in the indexes $ij$ which enter to Eq.~(\ref{eq:sysraw}). Basing
on Eq.~(\ref{eq:corrXi}), we rewrite Eq.~(\ref{eq:sysraw}) in
the form,
\begin{align}\label{eq:sysSA}
  \dot{S}(k,t)+2\eta_{m}k^{2}S(k,t) = &
  \frac{\tau_{d}}{2(P+\rho)} 
  \bigg\{
    -k^{2}S(k,t)\int\frac{\mathrm{d}^{3}q}{(2\pi)^{3}}S(q,t)
    \left[
      3-(\hat{\mathbf{k}}\hat{\mathbf{q}})^{2}
    \right]
    \notag
    \\
    & +
    kA(k,t)\int\frac{\mathrm{d}^{3}q}{(2\pi)^{3}}qA(q,t)
    \left[
      3-(\hat{\mathbf{k}}\hat{\mathbf{q}})^{2}
    \right]
  \bigg\},
  \nonumber
  \\
  \dot{A}(k,t)+2\eta_{m}k^{2}A(k,t) = & 
  \frac{\tau_{d}}{2(P+\rho)} 
  \bigg\{  
    -k^{2}A(k,t)\int\frac{\mathrm{d}^{3}q}{(2\pi)^{3}}S(q,t)
    \left[
      3-(\hat{\mathbf{k}}\hat{\mathbf{q}})^{2}
    \right]
    \notag
    \\
    & +
    kS(k,t)\int\frac{\mathrm{d}^{3}q}{(2\pi)^{3}}qA(q,t)
    \left[
      3-(\hat{\mathbf{k}}\hat{\mathbf{q}})^{2}
    \right]
  \bigg\}.
\end{align}
Using Eq.~(\ref{eq:SAdef}), we express Eq.~(\ref{eq:sysSA}) as
\begin{align}\label{eq:sysrhoh}
  \frac{\partial\rho(k,t)}{\partial t} & =-2\eta_{\mathrm{eff}}k^{2}\rho(k,t)+\alpha_{\mathrm{eff}}k^{2}h(k,t),
  \nonumber
  \\
  \frac{\partial h(k,t)}{\partial t} & =-2\eta_{\mathrm{eff}}k^{2}h(k,t)+4\alpha_{\mathrm{eff}}\rho(k,t),
\end{align}
where
\begin{equation}\label{eq:etaalpha}
  \eta_{\mathrm{eff}}=\eta_{m}+\frac{4}{3}\frac{\tau_{d}}{P+\rho}\int\mathrm{d}q\rho(q,t),
  \quad
  \alpha_{\mathrm{eff}}=\frac{2}{3}\frac{\tau_{d}}{P+\rho}\int\mathrm{d}qq^{2}h(q,t).
\end{equation}
Equations~(\ref{eq:sysrhoh}) and~(\ref{eq:etaalpha}) coincide
with those in Ref.~\refcite{Cam07}.

\section{New variables for the numerical simulation of the HMFs evolution\label{sec:NEWVAR}}

In the numerical solution of Eq.~(\ref{eq:HMFsys}), we use
the new variables~\cite{DvoSem21,Dvo22},
\begin{align}\label{eq:newvar}
  \tilde{\mathcal{E}}_{{\rm B_{\mathrm{Y}}}}(\tilde{k},\tilde{\eta}) & =
  \frac{\tilde{k}_{\mathrm{max}}\pi^{2}}{6\alpha'^{2}}R(\kappa,\tau),
  \quad
  \tilde{\mathcal{H}}_{{\rm B_{\mathrm{Y}}}}(\tilde{k},\tilde{\eta})=\frac{\pi^{2}}{3\alpha'^{2}}H(\kappa,\tau),
  \nonumber
  \\
  \xi_{\mathrm{R,L,0}}(\tilde{\eta}) & =
  \frac{\pi\tilde{k}_{\mathrm{max}}}{\alpha'}M_{\mathrm{R,L,0}}(\tau),
  \quad
  \tau=\frac{2\tilde{k}_{\mathrm{max}}^{2}}{\sigma_{c}}\tilde{\eta},
  \quad
  \tilde{k}=\tilde{k}_{\mathrm{max}}\kappa,
\end{align}
 where $0<\kappa<1$ and $\tau\geq0$ is the new dimensionless time.
Using Eq.~(\ref{eq:newvar}), we rewrite Eq.~(\ref{eq:HMFsys})
in the form~\cite{DvoSem21,Dvo22},
\begin{align}\label{eq:newsys}
  \frac{\partial R}{\partial\tau}= & -\kappa^{2}
  \left(
    1+\lambda_{t}I_{\mathrm{R}}\right)R+\kappa^{2}
    \left(
      M_{\mathrm{R}}-\frac{M_{\mathrm{L}}}{2}+\lambda_{t}I_{\mathrm{H}}
  \right)
  H,
  \nonumber
  \displaybreak[2]
  \\
  \frac{\partial H}{\partial\tau}= & -\kappa^{2}
  \left(
    1+\lambda_{t}I_{\mathrm{R}}
  \right)
  H+
  \left(
    M_{\mathrm{R}}-\frac{M_{\mathrm{L}}}{2}+\lambda_{t}I_{\mathrm{H}}
  \right)R,
  \nonumber
  \displaybreak[2]
  \\
  \frac{\mathrm{d}M_{\mathrm{R}}}{\mathrm{d}\tau}= &
  I_{\mathrm{H}}-\left(M_{\mathrm{R}}-\frac{M_{\mathrm{L}}}{2}\right)I_{\mathrm{R}}-
  \Gamma'(M_{\mathrm{R}}-M_{\mathrm{L}}+M_{0}),
  \nonumber
  \displaybreak[2]
  \\
  \frac{\mathrm{d}M_{\mathrm{L}}}{\mathrm{d}\tau}= &
  -\frac{1}{4}I_{\mathrm{H}}+\frac{1}{4}
  \left(
    M_{\mathrm{R}}-\frac{M_{\mathrm{L}}}{2}
  \right)
  I_{\mathrm{R}}-\Gamma'(M_{\mathrm{L}}-M_{\mathrm{R}}-M_{0})/2-\frac{\Gamma'_{s}}{2}M_{\mathrm{L}},
  \nonumber
  \displaybreak[2]
  \\
  \frac{\mathrm{d}M_{0}}{\mathrm{d}\tau}= & -\Gamma'(M_{\mathrm{R}}+M_{0}-M_{\mathrm{L}})/2,
\end{align}
where 
\begin{align}
  I_{\mathrm{R}}(\tau)= & \int_{0}^{1}\mathrm{d}\kappa R(\kappa,\tau),
  \quad
  I_{\mathrm{H}}(\tau)=\int_{0}^{1}\mathrm{d}\kappa \kappa^{2}H(\kappa,\tau),
  \quad
  \lambda_{t}=\frac{2\sigma_{c}\tilde{k}_{\mathrm{max}}^{2}\pi^{2}}{9\alpha'^{4}(\tilde{p}+\tilde{\rho})},
  \nonumber
  \\
  \Gamma'(\tau)= & \frac{121\sigma_{c}}{\tilde{\eta}_{\mathrm{EW}}\tilde{k}_{\mathrm{max}}^{2}}
  \left[
    1-\frac{T_{\mathrm{EW}}^{2}}{T_{\mathrm{RL}}^{2}}
    \left(
      1+\frac{T_{\mathrm{RL}}}{M_{0}}\frac{\sigma_{c}}{2\tilde{k}_{\mathrm{max}}^{2}}\tau
    \right)^{2}
  \right],
\end{align}
and $\Gamma'_{s}=\sigma_{c}\Gamma_{\mathrm{sph}}/2\tilde{k}_{\mathrm{max}}^{2}$.

\section{Energy spectrum of a stochastic GW background\label{sec:OMEGDER}}

It is convenient to characterize the spectrum of isotropic stochastic GWs by the following dimensionless function of the frequency $f$ measured in Hz~\cite{AllOtt97}:
\begin{equation}
  \Omega(f) = \frac{1}{\rho_\mathrm{crit}}\frac{\text{d}\rho_\mathrm{GW}}{\text{d}\ln f},
\end{equation}
where $\text{d}\rho_\mathrm{GW}$ is the energy density of GWs contained within the frequency interval $(f, f + \text{d}f)$ and $\rho_\mathrm{crit}$ is the critical density of the universe defined in Sec.~\ref{subsec:RESGWs}. Using the definition of the total conformal energy density of GWs in Eq.~\eqref{eq:densGW}, we get that 
\begin{equation}\label{eq:densGWf}
  \rho_{\mathrm{GW}}^{(c)}(\eta)=\int_{0}^{\infty}\rho_{\mathrm{GW}}^{(c)}(f,\eta)\mathrm{d}f =
  \int_{0}^{\infty} \frac{\text{d}\rho_\mathrm{GW}}{\text{d}f} \mathrm{d}f.
\end{equation}
On the basis of Eq.~\eqref{eq:densGWf}, one obtaines that $\tfrac{\text{d}\rho_\mathrm{GW}}{\text{d}f} = \rho_{\mathrm{GW}}^{(c)}(f,\eta)$.

Taking into account that $k=2\pi f$ for massless gravitons, we get that $\rho_{\mathrm{GW}}^{(c)}(f,\eta) = 2\pi \rho_{\mathrm{GW}}^{(c)}(k,\eta)$, where $\rho_{\mathrm{GW}}^{(c)}(k,\eta)$ is given in Eq.~\eqref{eq:rhoGWisotr}. Eventually, we arrive to Eq.~\eqref{eq:Omegadef} with $\rho_{\mathrm{GW}}(f) \equiv \rho_{\mathrm{GW}}^{(c)}(f,\eta)$. Since the frequency $f$ is related to the conformal momentum $k$, we call it the conformal frequency.

\section{Derivation of the density matrix equation for neutrino oscillations\label{sec:DENSMATR}}

In this appendix, we clarify some of the issues in the derivation
of the equation for the density matrix of flavor neutrinos interacting
with stochastic GWs. The treatment of these issues in Ref.~\refcite{Dvo21}
was insufficiently strict.

The evolution of the wavefunction of flavor neutrinos $\nu^{\mathrm{T}}=(\nu_{e},\nu_{\mu},\nu_{\tau})$
under the influence of stochastic GWs obeys the effective Schr\"odinger
equation, $\mathrm{i}\dot{\nu}=(H_{f}^{(0)}+H_{f}^{(1)})\nu$, where
$H_{f}^{(0,1)}$ are the effective Hamiltonians for neutrino oscillations
in vacuum and the contribution of stochastic GWs. The explicit form
of $H_{f}^{(0,1)}$ is given in Refs.~\refcite{Dvo19,Dvo20,Dvo21}.
We can also consider neutrino mass eigenstates $\psi_{a}$, $a=1,2,3$,
having the masses $m_{a}$, by making the matrix transformation, $\nu=U\psi$,
where $U$ is the $3\times3$ unitary matrix. The effective Hamiltonians
in the mass basis are $H_{m}^{(0,1)}=U^{\dagger}H_{f}^{(0,1)}U$.
It turns out that $H_{m}^{(0,1)}$ are diagonal, $\left(H_{m}^{(1,2)}\right)_{ab}\propto\delta_{ab}$.

We can also consider the wavefunction in the interaction picture $\psi'$,
$\psi=\exp(-\mathrm{i}H_{m}^{(0)}t)\psi'$. It obeys the Schr\"odinger
equation, $\mathrm{i}\dot{\psi}'=H_{\mathrm{int}}\psi',$ where $H_{\mathrm{int}}=\exp(\mathrm{i}H_{m}^{(0)}t)H_{m}^{(1)}\exp(-\mathrm{i}H_{m}^{(0)}t)=H_{m}^{(1)}$
is the Hamiltonian in the interaction picture. One can prove that
$H_{\mathrm{int}}=H_{m}^{(1)}$ (see, e.g., Ref.~\refcite{Dvo21}) since
both $H_{m}^{(0)}$ and $H_{m}^{(1)}$ are diagonal. If we consider
stochastic external fields, instead of $\psi'$, it is convenient
to deal with the density matrix $\rho'$ which satisfies the equation,
\begin{equation}\label{eq:rhoprimeeq}
  \mathrm{i}\dot{\rho}'=[H_{m}^{(1)},\rho'].
\end{equation}
The initial condition for Eq.~(\ref{eq:rhoprimeeq}) is $\rho'(0)=U^{\dagger}\rho_{f}^{(0)}U$,
where $\rho_{f}^{(0)}$ is the initial density matrix for flavor neutrinos.
We suppose that $\left(\rho_{f}^{(0)}\right)_{\lambda\lambda'}\propto\delta_{\lambda\lambda'}\left(F_{\lambda}\right)_{\mathrm{S}}$,
where $\left(F_{\lambda}\right)_{\mathrm{S}}$ are the fluxes of flavor
neutrinos at a source.

The formal solution of Eq.~(\ref{eq:rhoprimeeq}) is
\begin{equation}\label{eq:rhoformsol}
  \rho'(t)=\rho'(0)-\mathrm{i}
  \int_{0}^{t}[H_{m}^{(1)}(t_{1}),\rho'(0)]\mathrm{d}t_{1}-
  \int_{0}^{t}\int_{0}^{t_{1}}[H_{m}^{(1)}(t_{1}),[H_{m}^{(1)}(t_{2}),\rho'(0)]]\mathrm{d}t_{1}\mathrm{d}t_{2}+\dotsc,
\end{equation}
where $H_{m}^{(1)}=H_{m}^{(1)}(t)$ is supposed to be the random function
of time since we consider stochastic GWs which have randomly distributed
amplitudes and cross the neutrino trajectory at random angles. GW
can have two independent polarizations: `plus' and `times'. Thus,
we separate the effective Hamiltonian into two parts, $H_{m}^{(1)}=H_{m,+}^{(1)}+H_{m,\times}^{(1)}$,
which are uncorrelated. After averaging Eq.~(\ref{eq:rhoformsol}),
one gets
\begin{align}\label{eq:rhoav}
  \left\langle
    \rho'
  \right\rangle (t)= &
  \rho'(0)-\int_{0}^{t}\int_{0}^{t_{1}}
  \left[
    \left\langle
    H_{m,+}^{(1)}(t_{1}),
    \left[
    H_{m,+}^{(1)}(t_{2})
    \right\rangle,
    \rho'(0)
    \right]
  \right]
  \mathrm{d}t_{1}\mathrm{d}t_{2}
  \notag
  \\
  & -
  \int_{0}^{t}\int_{0}^{t_{1}}
  \left[
  \left\langle
  H_{m,\times}^{(1)}(t_{1}),
  \left[
  H_{m,\times}^{(1)}(t_{2})
  \right\rangle,
  \rho'(0)
  \right]
  \right]
  \mathrm{d}t_{1}\mathrm{d}t_{2}+\dotsc.
\end{align}
Let us consider, e.g., the term containing the `plus' correlator in
Eq.~(\ref{eq:rhoav}). Using the results of Ref.~\refcite{Dvo21},
we have
\begin{equation}
\left(H_{m,+}^{(1)}\right)_{ab}=-\frac{p^{2}}{2E_{a}}\delta_{ab}h_{+}\sin^{2}\vartheta\cos2\varphi\cos\phi_{a},\label{eq:Hplus}
\end{equation}
where $\phi_{a}=\omega t(1-v_{a}\cos\vartheta)$ is the phase of GW
accounting for the intersection of the direction of the GW propagation
and the neutrino trajectory, $v_{a}=p/E_{a}$ is the velocity of the
neutrino mass eigenstate, $E_{a}=\sqrt{m_{a}^{2}+p^{2}}$ is the energy
of the neutrino mass eigenstate, $p$ is the momentum of neutrinos,
$\vartheta$ and $\varphi$ are the angles fixing the neutrino momentum
with respect to the wave vector of GW, which is supposed to propagate
along the $z$-axis.

In the averaging procedure in Eq.~(\ref{eq:rhoav}), one has to deal
with the binary combination of the components of $H_{m,+}^{(1)}$.
Thus, accounting for Eq.~(\ref{eq:Hplus}), we should consider the
mean value
\begin{multline}\label{eq:avhang}
  \frac{p^{4}}{4E_{a}E_{b}}
  \left\langle
    h'_{+}h''_{+}\sin^{2}\vartheta'\sin^{2}\vartheta''\cos2\varphi'\cos2\varphi''\cos\phi'_{a}\cos\phi''_{b}
  \right\rangle 
  \\
  =
  \frac{p^{4}}{4E_{a}E_{b}}
  \left\langle
    h'_{+}h''_{+}
  \right\rangle
  \left\langle
    \sin^{2}\vartheta'\sin^{2}\vartheta''\cos2\varphi'\cos2\varphi''\cos\phi'_{a}\cos\phi''_{b}
  \right\rangle,
\end{multline}
where prime and double prime correspond to different moments of time
$t'$ and $t''$. Also, we separated the mean values of the amplitudes
and the angular factors. The correlator of the amplitudes $\left\langle h'_{+}h''_{+}\right\rangle =f_{+}(t'-t'')$
is the arbitrary function depending on the source of GWs. In Sec.~\ref{sec:NUFLOSC},
we relate it to the spectral density of GW; cf. Eq.~(\ref{eq:specdens}).

We assume that random angles $\vartheta$ and $\varphi$ are $\delta$-function
correlated. It is a reasonable assumption since sources of GWs are
distributed randomly. Thus, in Eq.~\eqref{eq:avhang}, we obtain that
\begin{multline}\label{eq:avang}
  \left\langle
    \sin^{2}\vartheta'\sin^{2}\vartheta''\cos2\varphi'\cos2\varphi''\cos\phi'_{a}\cos\phi''_{b}
  \right\rangle
  \to
  \left\langle
    \sin^{4}\vartheta\cos^{2}2\varphi\cos\phi{}_{a}\cos\phi_{b}
  \right\rangle 
  \\
  =
  \int_{0}^{2\pi}\frac{\mathrm{d}\varphi}{2\pi}\cos^{2}2\varphi
  \int_{0}^{\pi}\frac{\mathrm{d}\vartheta}{\pi}\sin^{4}\vartheta\cos[\omega t(1-v_{a}\cos\vartheta)]
  \cos[\omega t(1-v_{a}\cos\vartheta)].
\end{multline}
The integral over $\vartheta$ in Eq.~(\ref{eq:Iab}) reads
\begin{align}\label{eq:Iab}
  I_{ab}= & \frac{1}{2}\int_{0}^{\pi}\frac{\mathrm{d}\vartheta}{\pi}\sin^{4}\vartheta
  \left[
    \cos(\delta{}_{ab}\cos\vartheta)+\cos(2\omega t)\cos(\sigma{}_{ab}\cos\vartheta)+\sin(2\omega t)\sin(\sigma{}_{ab}\cos\vartheta)
  \right]
  \nonumber
  \\
  & =
  \frac{3}{16}
  \left[
    I_{c}(\delta{}_{ab})+\cos(2\omega t)I_{c}(\sigma{}_{ab})+\sin(2\omega t)I_{s}(\sigma{}_{ab})
  \right],
\end{align}
where $\delta{}_{ab}=(v_{a}-v_{b})\omega t$ and $\sigma{}_{ab}=(v_{a}+v_{b})\omega t$.
The integrals in Eq.~(\ref{eq:Iab}) are~\cite{Dvo19}
\begin{align}
  I_{c} & (\lambda)=\frac{8}{3\pi}\int_{0}^{\pi}\mathrm{d}\vartheta\sin^{4}\vartheta\cos(\lambda\cos\vartheta)=
  \frac{8}{\lambda^{3}}
  \left[
    2J_{1}(\lambda)-\lambda J_{0}(\lambda)
  \right],
  \nonumber
  \\
  I_{s} & (\lambda)=\frac{8}{3\pi}\int_{0}^{\pi}\mathrm{d}\vartheta\sin^{4}\vartheta\sin(\lambda\cos\vartheta)=0,
\end{align}
where $J_{0,1}(\lambda)$ are the Bessel functions.

In Sec.~\ref{subsec:RESGWs}, we find that the spectral function
$\Omega$ for relic GWs produced by random HMFs reaches its maximal
values in a plateau with the the frequencies range from $\sim\text{Hz}$
to $\sim\text{kHz}$ (see, e.g., Table~\ref{tab:plat}). Thus, if
study the interaction of SN neutrinos with such stochastic GWs, we
get that the quantities in Eq.~\ref{eq:Iab} are
\begin{align}\label{eq:sigmadelta}
  \min
  \left(
    \sigma_{ab}
  \right) & =
  2\omega_{\text{min}}L=1.3\times10^{13}\gg1,
  \nonumber
  \\
  \max
  \left(
    \delta_{ab}
  \right) & =
  \omega_{\text{max}}L\frac{\Delta m_{31}^{2}}{2E^{2}}=8.3\times10^{-2}\ll1,
\end{align}
where $L=10\,\text{kpc}$ is the neutrino propagation distance comparable
with the galaxy size (see Ref.~\refcite{Dvo21} and Sec.~\ref{sec:NUFLOSC})
and $E=10\,\text{MeV}$ is the typical SN neutrino energy. We take
$\Delta m_{31}^{2}=m_{3}^{2}-m_{1}^{2}=2.55\times10^{-3}\,\text{eV}^{2}$~\cite{Sal20}
in Eq.~(\ref{eq:sigmadelta}) since it is the maximal mass squared
difference.

Using the estimates in Eq.~(\ref{eq:sigmadelta}), one gets that
$I_{c}(\lambda)\to1$ if $\lambda\ll1$ and $I_{c}(\lambda)\to0$
if $\lambda\gg1$. Thus, $I_{ab}\to\tfrac{3}{16}$ in Eq.~(\ref{eq:Iab}).
It is important that the leading order in $I_{ab}$ is independent
of the indexes $a$ and $b$, as well as of time. It means that, effectively,
we can set $v_{a}\to1$ in $\phi_{a}$. The same result is obtained
while averaging $\left\langle H_{m,\times}^{(1)}(t_{1})H_{m,\times}^{(1)}(t_{2})\right\rangle $.
The present averaging over the angles $\vartheta$ and $\varphi$
is careful, whereas the analysis in Ref.~\refcite{Dvo21} was insufficiently
strict.

Finally, Eq.~(\ref{eq:rhoav}) is rewritten in the form,
\begin{align}\label{eq:rhoavfin}
  \left\langle
    \rho'
  \right\rangle (t)= &
  \rho'(0)-\frac{3}{128}[H_{m}^{(0)},[H_{m}^{(0)},\rho'(0)]]
  \notag
  \\
  & \times
  \int_{0}^{t}\int_{0}^{t_{1}}
  \left(
    \left\langle
      h_{+}(t_{1})h_{+}(t_{2})
    \right\rangle +
    \left\langle
      h_{\times}(t_{1})h_{\times}(t_{2})
    \right\rangle
  \right)
  \mathrm{d}t_{1}\mathrm{d}t_{2}+\dotsc,
\end{align}
where $H_{m}^{(0)}=\tfrac{1}{2E}\text{diag}\left(0,\Delta m_{21}^{2},\Delta m_{31}^{2}\right)$
is the effective Hamiltonian for vacuum oscillations. This series
in Eq.~(\ref{eq:rhoavfin}) is a formal solution of the following
differential equation:
\begin{align}\label{eq:rhoeq}
  \frac{\mathrm{d}}{\mathrm{d}t}
  \left\langle
    \rho'
  \right\rangle (t)= & -g[H_{m}^{(0)},[H_{m}^{(0)},
  \left\langle
    \rho'
  \right\rangle (t)]],
  \notag
  \\
  g(t) = & \frac{3}{128}\int_{0}^{t}
  \left(
    \left\langle
      h_{+}(t)h_{+}(t_{1})
    \right\rangle +
    \left\langle
      h_{\times}(t)h_{\times}(t_{1})
    \right\rangle
  \right)
  \mathrm{d}t_{1}.
\end{align}
Applying the formalism developed in Ref.~\refcite{Dvo21} to Eq.~(\ref{eq:rhoeq}),
we obtain the probabilities in Eq.~(\ref{eq:Plambdag}).

\end{document}